\RequirePackage{fix-cm}
\documentclass[%
 reprint,
 amsmath,amssymb,
 aps,
]{revtex4-2}

\usepackage{graphicx}
\usepackage{dcolumn}
\usepackage{bm}

\usepackage{amsfonts}
\usepackage{amsmath}
\usepackage{booktabs}
\usepackage{multirow}
\usepackage{placeins}
\usepackage{pgf}
\usepackage{nicefrac}
\usepackage{siunitx}
\usepackage{amssymb}
\usepackage{array}
\usepackage{graphicx}
\usepackage{dsfont}
\newcolumntype{C}[1]{>{\centering\arraybackslash}m{#1}}
\newcolumntype{R}[1]{>{\raggedleft\arraybackslash}m{#1}}
\usepackage{hyperref}
\usepackage{cleveref}
\hypersetup{colorlinks=true}

\definecolor{slateblue}{HTML}{31207F}
\definecolor{darkviolet}{HTML}{7B207F}
\definecolor{crimson}{HTML}{7F1132}

\hypersetup{
    colorlinks=true,
    linkcolor=slateblue,
    citecolor=darkviolet,
    urlcolor=crimson
}

\DeclareMathOperator{\ord}{\mathcal{O}}
\newcommand{\order}[1]{\ensuremath{\ord\left(#1\right)}}

\begin{document}

\preprint{APS/123-QED}

\title{Optimised 8th-Order Trotter-Suzuki Schemes\\for Quantum Simulation}

\author{Marko Maležič}
 \email{malezic@hiskp.uni-bonn.de}
\author{Johann Ostmeyer}%
 \email{ostmeyer@hiskp.uni-bonn.de}
\affiliation{%
 Helmholtz-Institut f\"ur Strahlen- und Kernphysik,\\University of Bonn, Nussallee 14-16,\\53115 Bonn, Germany
}%

\date{\today}

\begin{abstract}
  Simulating quantum dynamics in many-body systems is one of the most promising applications of quantum computers.
  Product formulas are a natural choice for this task, since they map directly onto quantum gates, and increasing their order $n$ could lead to reduced gate depths, provided the parameters are chosen efficiently.
  So far, schemes beyond $6^{\textrm{th}}$ order have only been built as compositions of second-order schemes, while more general product formulas are known only up to $6^{\textrm{th}}$ order.
  In this work, we extend our optimisation framework to order $n = 8$, and find both more efficient compositions and the first $8^{\textrm{th}}$ order general schemes.
  Our formulas improve on the theoretical efficiency of the state-of-the-art schemes by Morales \textit{et al}.~\cite{morales2022greatly} and we recommend two schemes explicitly.
  In simulations of the Heisenberg XZ chain one of our recommended schemes always outperforms those by Morales \textit{et al}, though it is not always the same scheme that performs best.
  More generally, we observe that for Hamiltonians split into many non-commuting operators the uniformity of a Trotter decomposition becomes the main indicator of a scheme's performance.
  To assess the schemes' use on quantum hardware, we simulate noisy circuits with our new Qiskit plugin \texttt{qiskit-omelyan}.
  At current noise levels, the Leapfrog scheme remains the best choice, but as noise rates decrease toward fault tolerance, our recommended schemes become competitive.
\end{abstract}

\maketitle



\section{Introduction}\label{sec:intro}

Calculating the time evolution of many-body systems is a highly non-trivial problem in both the classical and quantum world.
Quantum many-body systems feature a Hilbert space dimension which grows exponentially with the system size as $2^{L}$ for $L$ spins, and makes exact methods infeasible on classical hardware~\cite{lami2025beginnerslecturenotesquantum}.
Quantum computers reduce the exponential memory cost by storing the states in only a linear number of qubits, but require that time evolution be broken down into a sequence of unitary few-qubit gates~\cite{Lloyd1996}.
Such a decomposition is generally an approximation, and many methods have been proposed to accurately construct it while keeping the gate count small.
These include product formulas~\cite{trotter_original, Suzuki:1976be}, linear combinations of unitaries (LCU)~\cite{Childs:2012gwh}, quantum signal processing (QSP)~\cite{PhysRevLett.118.010501}, and randomised methods~\cite{Childs_2019,PhysRevLett.123.070503}.
For classical systems, symplectic integrators follow the exact dynamics of a slightly perturbed Hamiltonian, such that the energy error remains bounded instead of drifting~\cite{Hairer:2006gni}.
This stays true even for chaotic dynamics which amplify any integration error exponentially, so that no integrator can follow individual trajectories over long times.

Product formulas, also known as Trotter-Suzuki schemes, are especially favoured, because they do not require ancillas or block encodings, and each factor maps directly onto gates.
The time evolution operator $U$ for a two-part Hamiltonian $H = A + B$ can be approximated by a product

\begin{equation}
  U (t) = e^{-i t H} \approx \left[ S_{n} (h) \right]^{\nicefrac{t}{h}},\label{eq:evolve_op}
\end{equation}

\noindent
where $S_{n} (h)$ is a product formula of order $n$ at a step size $h$, composed of exponentials $e^{-i \mathsf{a}_{j} h A}$ and $e^{-i \mathsf{b}_{j} h B}$ with scheme parameters $\mathsf{a}_{j}$ and $\mathsf{b}_{j}$.
These factors are unitary, so the approximation is exactly unitary as well.
If $A$ and $B$ are each sums of commuting local terms, the factors also decompose exactly into few-qubit gates.
The same splitting applied to classical Hamiltonian flow yields symplectic integrators~\cite{Hairer:2006gni}, so the schemes apply equally to classical simulations.
The order $n$ corresponds to the scaling of the global error with the step size $\order{t h^{n}}$.
Although LCU and QSP achieve a better asymptotic scaling, product formulas remain competitive for systems of practical size, since their actual errors are much smaller than the theoretical bounds suggest~\cite{Childs:2018PNAS, PhysRevX.11.011020}.
This makes them a natural candidate for early fault-tolerant quantum computers~\cite{Campbell:2020wqh}.

In current quantum simulations, mostly the lowest orders $n = 1, 2$ are used, since they require the fewest gates per step.
However, for the global error to reach a target precision $\epsilon$ requires $\nicefrac{t}{h} \propto t^{1 + \nicefrac{1}{n}} \epsilon^{- \nicefrac{1}{n}}$ Trotter steps~\cite{PhysRevX.11.011020}.
By going to higher orders, the number of steps becomes all but linear in $t$, while the dependence on the precision becomes weak.
Fewer Trotter steps can lead to shallower circuits, which reduce the noise accumulation and allow for longer time evolution.
Higher-order formulas, on the other hand, consist of more exponentials per step, but their additional scheme parameters can be tuned to reduce the error prefactor, which determines their efficiency.

Such schemes are traditionally constructed as compositions of the second-order Leapfrog scheme, either recursively as proposed by Suzuki~\cite{SUZUKI1990319, Hatano_2005}, or numerically following Yoshida~\cite{YOSHIDA1990262}.
The state-of-the-art compositions are currently those by Morales \textit{et al}.~\cite{morales2022greatly} at orders $n = 8, 10$.
Omelyan \textit{et al}.~\cite{Omelyan_2002, OMELYAN2003272} instead derive the error terms of the product formula directly, which lifts the restriction to compositions and allows for more efficient schemes.
Ever better generic schemes have been found at orders $n = 4, 6$ initially by Omelyan \textit{et al}. themselves, later by Blanes and Moan~\cite{BLANES2002313}, and recently in our previous work~\cite{Malezic_2026}.
Beyond $6^{\textrm{th}}$ order, however, the number of error terms that have to be eliminated grows rapidly, and generic schemes have not been constructed so far.
For a broader overview we refer to the recent review~\cite{Ostmeyer:2026ckx} on Trotterization and other time evolution methods.

In this paper, we extend our framework based on Omelyan's method to order $n = 8$, find novel compositions as well as the first generic schemes at this order, and benchmark them in noiseless and noisy quantum simulations.
\Cref{sec:framework_extension} describes the $8^{\textrm{th}}$ order extension, the structure of its solutions, and how we have improved our minimisation routine to explore them.
There, we also discuss how our framework could be adapted to compositions at orders $n \geq 8$.
In \Cref{sec:results_order8}, we compare the theoretical efficiency of our schemes to the ones by Morales \textit{et al}.~\cite{morales2022greatly}.
We also perform numerical experiments on the Heisenberg XZ spin chain with two operator orderings, and study the error scaling with computational cost.
Tables~\ref{tab:scheme_q21} and \ref{tab:scheme_q30} provide the scheme parameters for highly efficient composition and generic schemes, respectively, and \Cref{app:additional_schemes} gives further schemes.
Finally, we simulate noisy circuits in \Cref{sec:quantum_circuits} with our newly developed plugin \texttt{qiskit-omelyan}~\cite{qiskit-omelyan} to assess when high-order product formulas could improve quantum simulations.
We conclude in \Cref{sec:conclusion}, where we also discuss ideas for future work.


\section{Framework extension to $\textrm{8}^{\textrm{th}}$ order}\label{sec:framework_extension}

Product formulas of order $n \geq 8$ have so far been constructed as compositions of lower-order schemes, either recursively, as proposed by Suzuki~\cite{SUZUKI1990319, Hatano_2005}, or with numerical optimisation like Yoshida's method~\cite{YOSHIDA1990262}.
The alternative approach by Omelyan \textit{et al}.~\cite{OMELYAN2003272}, which solves the order conditions of the splitting coefficients directly, has until now only been used up to order $n = 6$.
In previous work~\cite{Malezic_2026, Malezic:2026xyi}, we built a framework for constructing optimised schemes based on this approach.
Here, we extend the framework to order $n = 8$, discuss the manifold structures and explain some additions to our minimisation routine.

\subsection{The error term construction}\label{sec:term_construction}

To construct a scheme of arbitrary order from scratch, we expand it using the Baker--Campbell--Hausdorff (BCH) formula.
For the derivation, it suffices to assume a Hamiltonian composed of two non-commuting operators $H = A + B$, since a generalisation to $\Lambda > 2$ stages is readily available in~\cite{Ostmeyer:2022}.
We therefore make the ansatz

\begin{equation}
	\begin{split}
		&\exp \left[ (A + B) h + \sum\nolimits_{n} O_{n-1} h^{n-1} \right] = \\
		&= e^{\mathsf{a}_{1} h A} e^{\mathsf{b}_{1} h B} e^{\mathsf{a}_{2} h A} \cdots e^{\mathsf{b}_{q} h B} e^{\mathsf{a}_{q+1} h A} = \\
    &= e^{c_{1} h A} e^{c_{1} h B} e^{d_{1} h B} e^{d_{1} h A} \cdots e^{c_{q} h A} e^{c_{q} h B} e^{d_{q} h B} e^{d_{q} h A},
	\end{split}\label{eq:ansatz_BCH}
\end{equation}

\noindent
where $\mathsf{a}_{j}$ and $\mathsf{b}_{j}$ are the scheme parameters in Omelyan's notation, $h$ is the time step and $O_{n-1}$ denote the error terms.
Each sequence $e^{\mathsf{a}_{j} h A} e^{\mathsf{b}_{j} h B}e^{\mathsf{a}_{j+1} h A}$ forms one cycle, such that the scheme consists of $q$ (overlapping) cycles.
Furthermore, we assume a symmetric product formula, $\mathsf{a}_{j} = \mathsf{a}_{q+2-j}$ and $\mathsf{b}_{j} = \mathsf{b}_{q+1-j}$, which leaves $q + 1$ free parameters.
Then only odd powers of $h$ appear in the exponent, such that the sum runs over even $n = 2, 4, 6, \ldots$ only, and the schemes are of even order.
We also include the product formula in the ramp approach with parameters $c_{i}$, $d_{i}$, with the transformation between the two notations available in Ref.~\cite{Ostmeyer:2022}.
The error terms $O_{n-1}$ are composed of coefficients multiplied by commutators, which arise naturally from the BCH expansion.
The only exception is

\begin{equation}
  O_{1} = (\nu - 1) A + (\sigma - 1) B,\label{eq:O1_term}
\end{equation}

\noindent
where $\nu = \sum_{j} \mathsf{a}_{j} = 1$ and $\sigma = \sum_{j} \mathsf{b}_{j} = 1$ (or equivalently $\sum_{j} (c_{j} + d_{j}) = 1$) provide the first two constraints.

The structure of the error terms is the foundation of the framework, and we have previously derived them up to $O_{7}$ in Ref.~\cite{Malezic_2026}.
We now outline the derivation of the $O_{9}$ error term, since the full calculation is lengthy.
It can be found in our Mathematica~\cite{Mathematica} notebook (\texttt{Efficient\_Trotterization.nb}) in the repository~\cite{markomalezic_2026_18347430}, with the results collected in the notes (\texttt{Recursive\_formulae.pdf}) therein.
To make the derivation freely accessible, we also provide a port to Python as the package \texttt{pybch}, based on the open-source library SymPy~\cite{sympy}.

At order $n = 8$, the error term $O_{9}$ lives in the degree-9 component of the free Lie algebra generated by $A$ and $B$.
After removing linear dependencies due to antisymmetry and the Jacobi identity, $56$ independent commutators remain.
As in our previous work, we follow the right-nested commutator basis discussed by Arnal \textit{et al}.~\cite{Arnal:2020xpt}, ordered symmetrically under the exchange $A \leftrightarrow B$, such that $F_{57 - k}$ is obtained from $F_{k}$ by swapping $A$ and $B$.
The basis splits into 8 subspaces according to the number of $A$- and $B$-terms with dimensions $1, 4, 9, 14, 14, 9, 4, 1$, respectively.
We present one element from each subspace

\begin{equation}
  \begin{array}{r@{\;}c@{\;}l}
    F_{1} = [ A^{8} B ] & \longleftrightarrow & [ B^{8} A ] = F_{56}, \\[1pt]
    F_{2} = [ A^{3} B A^{4} B ] & \longleftrightarrow & [ B^{3} A B^{4} A ] = F_{55}, \\[1pt]
    & \vdots & \\[1pt]
    F_{6} = [ A^{3} B^{2} A^{3} B ] & \longleftrightarrow & [ B^{3} A^{2} B^{3} A ] = F_{51}, \\[1pt]
    & \vdots & \\[1pt]
    F_{15} = [ A^{2} B A^{2} B^{2} A B ] & \longleftrightarrow & [ B^{2} A B^{2} A^{2} B A ] = F_{42}, \\[1pt]
    & \vdots &
  \end{array}\label{eq:basis_n8}
\end{equation}

\noindent
where, for readability, we write the commutators in the compact right-nested form, e.g.\ $[A, [A, B]] = [A^{2} B]$.

With the commutator basis fixed, the error term can now be expanded in the basis as

\begin{equation}
  O_{9} = \sum_{k = 1}^{56} \epsilon_{k} F_{k},\label{eq:error_term_O9}
\end{equation}

\noindent
where the coefficients $\epsilon_{k}$ are polynomials in the scheme parameters $\mathsf{a}_{j}$ and $\mathsf{b}_{j}$.
Following the approach outlined in Sec.~2.2 of Ref.~\cite{Malezic_2026}, we derive recursive formulae in the number of cycles $q$ that express the $\epsilon_{k}$ in terms of the scheme parameters.
The recursion takes turns applying a step with the operator $A$ and operator $B$, and how many steps were taken is indicated by the indices $i_{A}$, $i_{B}$ in the expression for $\epsilon_{k, A}^{(i_{A}, i_{B})}$.
The simplest formula at $k = 56$ adds only one term for a step in $A$,

\begin{equation}
  \epsilon_{56, A}^{(i, i-1)} = \epsilon_{56, B}^{(i-1, i-1)} + \tilde{\epsilon}_{56} a_{i} \left( \sigma^{(i-1)} \right)^{8}.\label{eq:eps56}
\end{equation}

\noindent
where we express the scheme parameters $a_{i}, b_{i}$ in our notation from the previous paper~\cite{Malezic_2026}, which can be related to the notation by Omelyan.

The fractional prefactors $\tilde{\epsilon}_{k}$ fall out of the BCH expansion.
For example, the prefactor of the last commutator $F_{56}$ is

\begin{equation}
	\tilde{\epsilon}_{56} = -\frac{1}{1209600}
\end{equation}

\noindent
The rest of the basis, along with its prefactors, can be found in the notes in our repository~\cite{markomalezic_2026_18347430}.

The coefficient $\sigma^{(i-1)} = \sum\nolimits_{j}^{i-1} b_{j}$ sums up the parameters $b_{j}$ up to the recursion step $i-1$, which reaches the final form $\sigma$ as found in $O_{1}$~\eqref{eq:O1_term}.
Other expressions $\epsilon_{i, A}^{(i, i-1)}$ contain more terms and higher order coefficients.
These become lengthy, and we refer to our notes~\cite{markomalezic_2026_18347430} for details.

Combining the basis and the coefficients yields the error term $O_{9}$ in equation~\eqref{eq:error_term_O9}, which allows us to define the error function of order $n = 8$ as the Euclidean norm of its coefficients,

\begin{equation}
  \textrm{Err}_{8} = \sqrt{\sum\nolimits_{k=1}^{56} \left| \epsilon_{k} \right|^{2}}.\label{eq:err_n8}
\end{equation}

\noindent
The construction of $8^{\textrm{th}}$ order schemes thus becomes an optimisation problem: minimising $\textrm{Err}_{8}$ subject to the order conditions $O_{1} = O_{3} = O_{5} = O_{7} = 0$.
Before optimising, however, we must understand which parameter regions admit solutions at all, which requires a closer look at the structure of the solution manifold.

\begin{table}[t]
  \centering
  \caption{Order-condition counts for symmetric schemes of order $n \leq 12$.
           At order $n$, the coefficients of $O_{n-1}$ must vanish in addition to those of lower orders, which together give the number of constraints $T_{S}$ of general symmetric schemes.
           For symmetric compositions~\eqref{eq:composition}, these reduce to $T_{SS}$ conditions~\cite{McLachlan:1995otni}.
           The coefficients of the leading error term $O_{n+1}$ are hence found in the next row.
           The cycle range of each order starts at the minimal composition, $q = 2 T_{SS} - 1$, and ends before that of the next order.
           Generic solutions require $q \geq T_{S} - 1$, which for $n \geq 10$ (grey) lies beyond the cycle range, leaving only compositions.
           The coefficients for $n \leq 10$ agree with Ref.~\cite{Arnal:2020xpt}, and we extend them to $n = 12$.}\label{tab:order_properties}
  \begin{tabular*}{\columnwidth}{@{\extracolsep{\fill}}cccccc}
    \toprule
    Order $n$ & Coeff. & $T_{S}$ & $T_{SS}$ & Cycles $q$ & Generic $q$ \\
    \midrule
    2  & 2   & 2   & 1  & 1--2    & $\geq 1$  \\
    4  & 2   & 4   & 2  & 3--6    & $\geq 3$  \\
    6  & 6   & 10  & 4  & 7--14   & $\geq 9$  \\
    8  & 18  & 28  & 8  & 15--30  & $\geq 27$ \\
    10 & 56  & 84  & 16 & 31--66  & \textcolor{gray}{$\geq 83$}  \\
    12 & 186 & 270 & 34 & 67--146 & \textcolor{gray}{$\geq 269$} \\
    \bottomrule
  \end{tabular*}
\end{table}

\subsection{Order $\mathbf{n = 8}$ manifold structure}\label{sec:manifold_structure}

\begin{table*}[t]
	\centering
	\small
	\setlength{\tabcolsep}{5pt}
	\caption{Dimensions of the $8^{\textrm{th}}$ order solution manifolds, given by the number of null directions of the Jacobian $J_{ij}$.
         In the composition region, only compositions~\eqref{eq:composition} exist, and their submanifold gains a free direction with every second cycle $q$, as each pair of cycles adds one parameter.
         In the generic region, solutions also exist away from the compositions, and their manifold gains a free direction with every cycle.
         The composition submanifold keeps growing alongside, and remains of higher dimension throughout.}\label{tab:order8_regions}
  \begin{tabular}{l*{6}{C{1.05cm}}|*{4}{C{1.05cm}}}
    \toprule
    Region           & \multicolumn{6}{c|}{Composition region} & \multicolumn{4}{c}{Generic region} \\
    \midrule
    No.\ cycles $q$  & 15, 16 & 17, 18 & 19, 20 & 21, 22 & 23, 24 & 25, 26 & 27 & 28 & 29 & 30  \\
    Composition dim. & 0      & 1      & 2      & 3      & 4      & 5      & 6  & 6  & 7  & 7   \\
    Generic dim.     & --     & --     & --     & --     & --     & --     & 0  & 1  & 2  & 3   \\
    \bottomrule
  \end{tabular}
\end{table*}

To construct schemes of $8^{\textrm{th}}$ order within our framework, a total of 28 order constraints need to be satisfied, which generically requires 28 scheme parameters.
The total number of constraints $T_{S}$ at any order follows from Witt's formula, as demonstrated by McLachlan~\cite{McLachlan:1995otni}, and we list it for orders up to $n = 12$ in \Cref{tab:order_properties}.
Since a scheme with $q$ cycles has $q + 1$ parameters, generic solutions first appear at $q = T_{S} - 1$, which means $q = 27$ at $8^{\textrm{th}}$ order.

However, $8^{\textrm{th}}$ order schemes are known to exist at far fewer cycles, starting from $q = 15$.
These are Yoshida-type~\cite{YOSHIDA1990262} compositions, which chain $q$ lower-order formulas $S_{2}$ of different step sizes into a palindromic product,

\begin{equation}
  S_{n} (h) = \prod_{j=1}^{q} S_{2}(w_{j} h), \qquad w_{j} = w_{q+1-j},\label{eq:composition}
\end{equation}

\noindent
where we define the standard Leapfrog formula as,

\begin{equation}
  S_{2} (h) = e^{\frac{1}{2} h A} e^{h B} e^{\frac{1}{2} h A}
\end{equation}

\noindent
At $8^{\textrm{th}}$ order, such schemes were first constructed by Yoshida~\cite{YOSHIDA1990262}, and later improved by McLachlan~\cite{McLachlan:1995otni}, Kahan and Li~\cite{Kahan_Li_1997}, Sofroniou and Spaletta~\cite{Sofroniou_Spaletta_2005}, and most recently by Morales \textit{et al.}~\cite{morales2022greatly}.
Such compositions are naturally contained in our framework.
In the ramp notation~\eqref{eq:ansatz_BCH} compositions correspond to $c_{j} = d_{j}$, which combined with the palindromic structure results in the mirror symmetry $c_{j} = c_{q+1-j}$ (see Tab.~\ref{tab:scheme_q21}).
Restricted to compositions, the order constraints are no longer independent.
McLachlan~\cite{McLachlan:1995otni} showed that they reduce to $T_{SS}$ conditions, which we also list in \Cref{tab:order_properties}.
At $8^{\textrm{th}}$ order, only $T_{SS} = 8$ conditions remain, and since only every second step of the palindromic product adds a new parameter, compositions require $q \geq 2 T_{SS} - 1 = 15$.

The order $n = 8$ cycle range therefore splits into two regions.
Below $q = 27$ cycles, the parameters are too few to satisfy all constraints generically, and only $S_2$-compositions exist, which is why we refer to it as the composition region.
From 27 cycles onwards, solutions also exist separately from the compositions, which now form a lower-dimensional subset, and we refer to it as the generic region.
At higher orders, the generic region disappears, since $T_{S}$ grows much faster than $T_{SS}$, and generic solutions would only appear once compositions of the next order are already possible (see Tab.~\ref{tab:order_properties}).
In the following, we confirm this picture numerically at $8^{\textrm{th}}$ order.

The counting above also predicts the dimensions of the solution manifolds.
On the composition submanifold, a new free direction appears with every second cycle, starting from $q = 15$, while in the generic region, a new direction appears with every cycle, starting from $q = 27$.
To verify this, we examine the Jacobian

\begin{equation}
  J_{i j} = \frac{\partial C_{i}}{\partial c_{j}},
\end{equation}

\noindent
where $C_{i}$ denote the coefficients of the constrained error terms $O_{1}, \ldots, O_{7}$.
The number of vanishing singular values of $J_{i j}$ gives the local dimension of the solution manifold, i.e.\ the number of null directions along which we can optimise.
Sweeping through the number of cycles $q$, we evaluate the Jacobian at non-composition points, which yields the dimension of the generic manifold.
In composition solutions, we restrict it to schemes that preserve the mirror symmetry $c_{j} = c_{q+1-j}$, which yields the dimension of the composition submanifold.
We summarise our findings in \Cref{tab:order8_regions} and the analysis can be found in our Mathematica notebook (\texttt{Jacobian\_on\_manifold.nb}) in the repository~\cite{markomalezic_2026_18347430}.

\begin{figure*}[t]
	\centering
	\includegraphics[width=\textwidth]{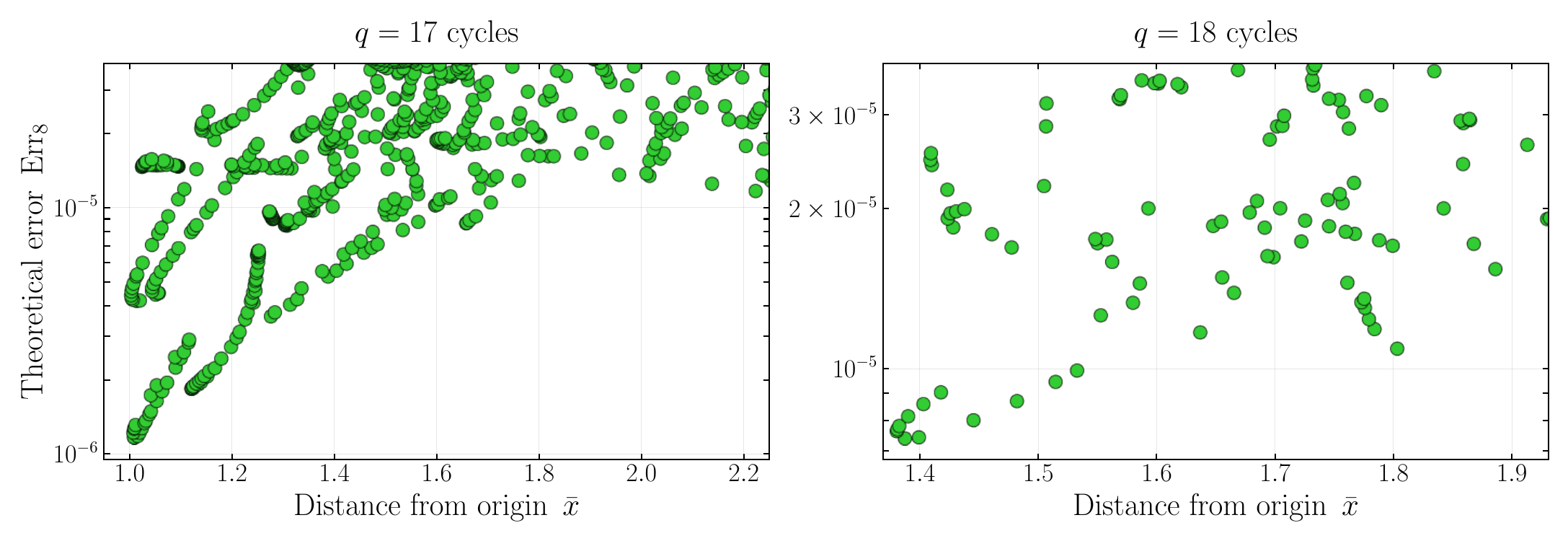}
	\caption{Solutions of the $8^{\textrm{th}}$ order conditions at $q = 17$ (left) and $q = 18$ cycles (right), classified by their theoretical error $\textrm{Err}_{8}$~\eqref{eq:err_n8} and distance $\bar{x}$~\eqref{eq:origin} from the point of equal scheme parameters.
         The solutions lie on curves, which trace out the one-dimensional composition submanifolds (see \Cref{tab:order8_regions}).}\label{fig:manifolds_q17-18}
\end{figure*}

As predicted, the dimension of the composition submanifold rises by one with every odd $q$, while the generic manifold starts from zero at $q = 27$ and rises with every cycle.
Since the mirror symmetry renders most constraints redundant, the composition submanifold remains of higher dimension throughout the generic region.
Both manifolds thus coexist as separate components of the solution set, which meet at most in isolated points.
In practice, this makes compositions hard to reach in the generic region, since random initial parameters break the mirror symmetry, and the minimiser converges to generic solutions instead.
In the composition region, on the other hand, no other solutions exist, and every converged run results in a composition.
We therefore adapt our framework to impose the mirror symmetry from the start, which gives access to the compositions in the generic region.
As we show in \Cref{sec:optim_improve}, it also improves the conditioning of the search in the composition region.

The one-dimensional composition submanifolds at $q = 17$ and $18$ can also be observed directly.
If we solve only the order conditions without optimising $\textrm{Err}_{8}$, the solutions land at arbitrary points on the submanifold.
We classify them by their theoretical error $\textrm{Err}_{8}$ and their distance $\bar{x}$ from the point of equal scheme parameters $c_{j} = \nicefrac{1}{2q}$,

\begin{equation}
  \bar{x}^{2} = 2 \sum_{j=1}^{q} \left( c_{j} - \frac{1}{2q} \right)^{2},\label{eq:origin}
\end{equation}

\noindent
where the factor $2$ accounts for the ramp-down parameters $d_{j} = c_{q+1-j}$.
While equation~\eqref{eq:origin} defines a 2-norm distance, a 1-norm definition is also possible, but we do not find a significant difference between the two.
As shown in \Cref{fig:manifolds_q17-18}, the solutions trace out distinct curves, as expected from a single free direction.
At $q = 18$, fewer solutions were found, which makes the curves harder to distinguish.
In fact, these curves were our first indication of the composition submanifold.
Optimising $\textrm{Err}_{8}$ along the free direction then leads to the minimum of each curve.
At more cycles $q\ge 19$, the submanifolds are of higher dimension, and their projections onto these two classifiers no longer form distinct curves.
In the following, we explain how our optimisation routines had to be adapted to efficiently minimise such curves at order $n = 8$.

\subsection{Optimisation improvements}\label{sec:optim_improve}

\begin{figure}[h!]
	\centering
	\includegraphics[width=\columnwidth]{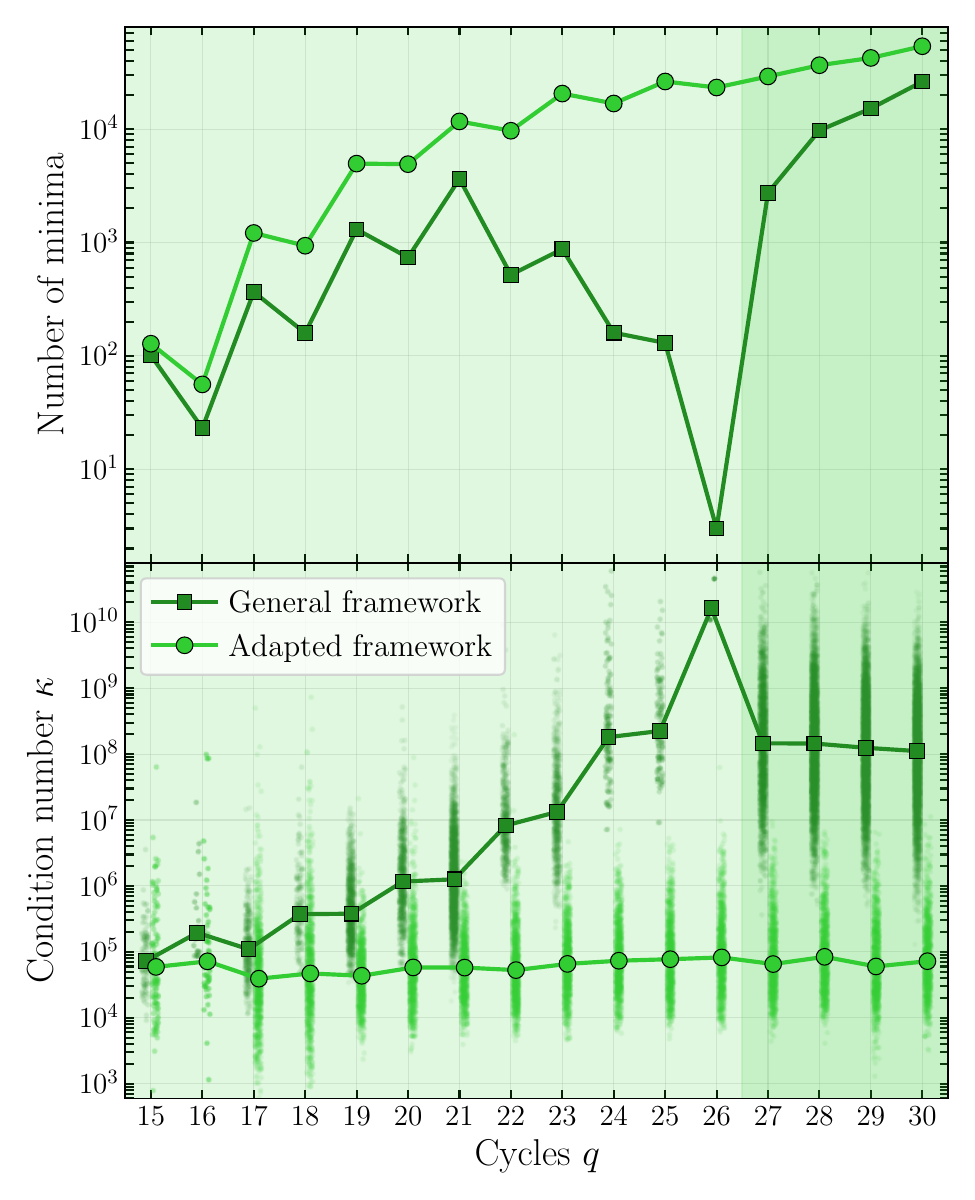}
	\caption{Number of distinct minima (top) and condition number $\kappa$~\eqref{eq:condition} of the Jacobian in the found solutions and their median value (bottom) for each number of cycles $q$ at order $n = 8$.
           The composition and generic regions (see Tab.~\ref{tab:order8_regions}) are shaded in light and dark green, respectively.
           The plots compare the results of the general framework and the composition adaptation with the imposed mirror symmetry $c_{j} = c_{q+1-j}$.
           Our general optimiser struggles to find composition schemes at higher cycles, as the condition number of the Jacobian blows up.
           The adaptation fixes this problem by restricting the search to composition schemes with fewer parameters and order conditions, which stabilises the conditioning and increases the minima count.
           Note that at higher cycles, the number of minima is almost certainly just a lower bound, and that only a subset of these minima was analysed further to compute the condition number.}
\label{fig:count_condition}
\end{figure}

We formulate our minimisation problem by first defining the $\chi^{2}$ function as

\begin{equation}
  \chi^{2} (\mathbf{p}) = \sum_{n\ge 2} \omega_{n} \textrm{Err}_{n}^{2} (\mathbf{p}) = \mathbf{y} (\mathbf{p})^{\dagger} \mathbf{W} \mathbf{y} (\mathbf{p}),\label{eq:min_chi2}
\end{equation}

\noindent
where $\textrm{Err}_{n}$ with $n < 8$ correspond to the order constraints.
The vector $\mathbf{p}$ stores the free scheme parameters, $\mathbf{y}$ the coefficients of $O_{3}, \ldots, O_{9}$, and the diagonal matrix $\mathbf{W}$ the weights $\omega_{n}$~\cite{Malezic_2026}.
As before, we minimise $\chi^{2}$ with the Levenberg--Marquardt algorithm~\cite{Levenberg:1944, Marquardt:1963}, which interpolates between the Gauss--Newton method and gradient descent through a damping parameter $\lambda$.
Each step $\mathbf{h}_{\textrm{LM}}$ is obtained by solving

\begin{equation}
  \left[ \mathbf{J}^{\dagger} \mathbf{W} \mathbf{J} + \lambda\, \textrm{diag}(\mathbf{J}^{\dagger} \mathbf{W} \mathbf{J}) \right] \mathbf{h}_{\textrm{LM}} = - \mathbf{J}^{\dagger} \mathbf{W} \mathbf{y} (\mathbf{p}),\label{eq:LM_step}
\end{equation}

\noindent
where $\mathbf{J} = \partial \mathbf{y} / \partial \mathbf{p}$ now also includes the coefficients of the leading error term.
As seen in \Cref{fig:count_condition}, the Jacobian becomes increasingly ill-conditioned with the number of cycles, as we show below.
The damping term keeps the matrix in Eq.~\eqref{eq:LM_step} invertible, as long as $\lambda$ is bounded from below by $\lambda_{\textrm{min}}$.
At order $n = 8$, we lowered this bound to $\lambda_{\textrm{min}} = 10^{-12}$, which still regularises the system, but damps the steps less strongly than before.

The second challenge is the computational cost, since both the number of parameters and the number of error coefficients grow, and each step becomes more expensive.
As the Levenberg--Marquardt algorithm is a local method, we must run many initial conditions to find a large set of minima, so each of them has to converge quickly.
As in Ref.~\cite{Malezic_2026}, we minimise in two stages.
We first minimise the full $\chi^{2}$ to reach the region of efficient schemes, and then impose the order constraints exactly by minimising only the constrained terms, starting from the result of the first stage.
For the first stage, we find the following weights to work well,

\begin{equation}
  \omega_{2} = 1, \;\; \omega_{4} = 10^{4}, \;\; \omega_{6} = 10^{7}, \;\; \omega_{8} = \left[ 10^{1}, 10^{3} \right],\label{eq:weights}
\end{equation}

\noindent
where we provide a range for $\omega_{8}$, since a smaller weight converges faster, but is likely to produce a less efficient scheme.
At orders $n \leq 6$, we froze the approximate Hessian $\mathbf{J}^{\dagger} \mathbf{W} \mathbf{J}$ of the first stage during the second, which keeps the solution close to the efficient region.
At order $n = 8$, however, this hampers the convergence of the second stage, so we recompute the Hessian instead, which increases the number of converged initial conditions.

The final improvement to convergence in our framework is the addition of geodesic acceleration~\cite{transtrum2012improvementslevenbergmarquardtalgorithmnonlinear}.
Interpreting the step $\mathbf{h}_{\textrm{LM}}$ as the velocity $\mathbf{v}$ along a geodesic path, we add a second-order correction from the acceleration $\mathbf{a}$, which solves

\begin{equation}
  \left[ \mathbf{J}^{\dagger} \mathbf{W} \mathbf{J} + \lambda\, \textrm{diag}(\mathbf{J}^{\dagger} \mathbf{W} \mathbf{J}) \right] \mathbf{a} = - \mathbf{J}^{\dagger} \mathbf{W} \mathbf{f}_{vv},\label{eq:geodesic_accel}
\end{equation}

\noindent
with the same matrix as in Eq.~\eqref{eq:LM_step}.
The improved step $\mathbf{h}_{\textrm{geod.}} = \mathbf{v} + \frac{1}{2} \mathbf{a}$ is only accepted if $2 \lvert \mathbf{a} \rvert / \lvert \mathbf{v} \rvert \leq 0.75$, which prevents the acceleration from dominating the step.
The directional second derivative $\mathbf{f}_{vv}$ is approximated by a forward finite difference along the normalised direction $\hat{\mathbf{v}} = \mathbf{v} / \lvert \mathbf{v} \rvert$,

\begin{equation}
  \mathbf{f}_{vv} \approx \frac{2 \lvert \mathbf{v} \rvert^{2}}{\delta} \left[ \frac{\mathbf{y}(\mathbf{p} + \delta\, \hat{\mathbf{v}}) - \mathbf{y}(\mathbf{p})}{\delta} - \mathbf{J} \hat{\mathbf{v}} \right],\label{eq:dir_deriv}
\end{equation}

\noindent
which requires only one additional evaluation, since $\mathbf{y}(\mathbf{p})$ and $\mathbf{J}$ have already been computed.
The normalisation keeps the perturbation well resolved even when the step becomes small close to convergence.
We use the same finite step $\delta$ as for the Jacobian, chosen as $\delta = \epsilon_{\textrm{mach}}^{1/3}$ to balance truncation and round-off errors.
The addition of geodesic acceleration moderately improves convergence, and it helped find some of the most efficient schemes at this order.

These improvements, however, work well only at smaller numbers of cycles.
At larger $q$, the unrestricted minimiser struggles to find solutions, especially in the composition region, where the count in \Cref{fig:count_condition} peaks at $q = 21$, before falling rapidly to only three schemes at $q = 26$.
We investigated this decline by computing the condition number $\kappa$ of the Jacobian $\mathbf{J}$

\begin{equation}
  \kappa = \frac{s_{\textrm{max}}}{s_{\textrm{min}}},\label{eq:condition}
\end{equation}

\noindent
which is the ratio between its maximal $s_{\textrm{max}}$ and minimal non-zero singular value $s_{\textrm{min}}$.
The bottom panel of \Cref{fig:count_condition} shows that the Jacobian becomes progressively ill-conditioned, with a peak at $q = 26$.
Since the minimiser operates on all $q$ scheme parameters, it has to suppress every direction that breaks the mirror symmetry, while converging onto a set where the Jacobian is rank-deficient.
In the generic region, no such symmetry has to be enforced, so the condition number plateaus and the count increases, since the minimiser can converge anywhere on the generic manifold.

We sought to remedy the conditioning problem by imposing the mirror symmetry $c_{j} = c_{q+1-j}$ discussed in \Cref{sec:manifold_structure}.
This adaptation to our framework roughly halves the number of parameters, and restricts our search to composition schemes only.
Removing the symmetry-breaking directions should considerably improve the conditioning of the Jacobian.
Indeed, with the adapted framework the median condition number remains at $\kappa \sim 10^{4}$--$10^{5}$ for all $q$, up to five orders of magnitude below the general framework, and the number of minima increases at every odd $q$.

The adaptation to compositions allowed us to explore the composition submanifold more thoroughly.
It yields new schemes at larger numbers of cycles in the composition region, and gives access to the compositions that persist in the generic region.
We discuss these results in \Cref{sec:results_order8}.
At order $n = 10$, generic schemes only appear at $q \geq 83$ (see \Cref{tab:order_properties}), so that within practical numbers of cycles, only compositions can be found, and this adaptation becomes essential.
The main effort there, however, lies in deriving the error terms, which we leave for future work.


\section{$\mathbf{8^{\textrm{th}}}$ order Trotterization results}\label{sec:results_order8}

In this section, we present the $8^{\textrm{th}}$ order schemes found by our minimiser, starting with their theoretical efficiency.
We then validate it with numerical experiments on the Heisenberg model, and conclude with our recommended scheme parameters at $n = 8$.

\subsection{Theoretical efficiency}\label{sec:theo_efficiency}

\begin{figure*}[t]
	\centering
	\resizebox{0.995\textwidth}{!}{%
		\input{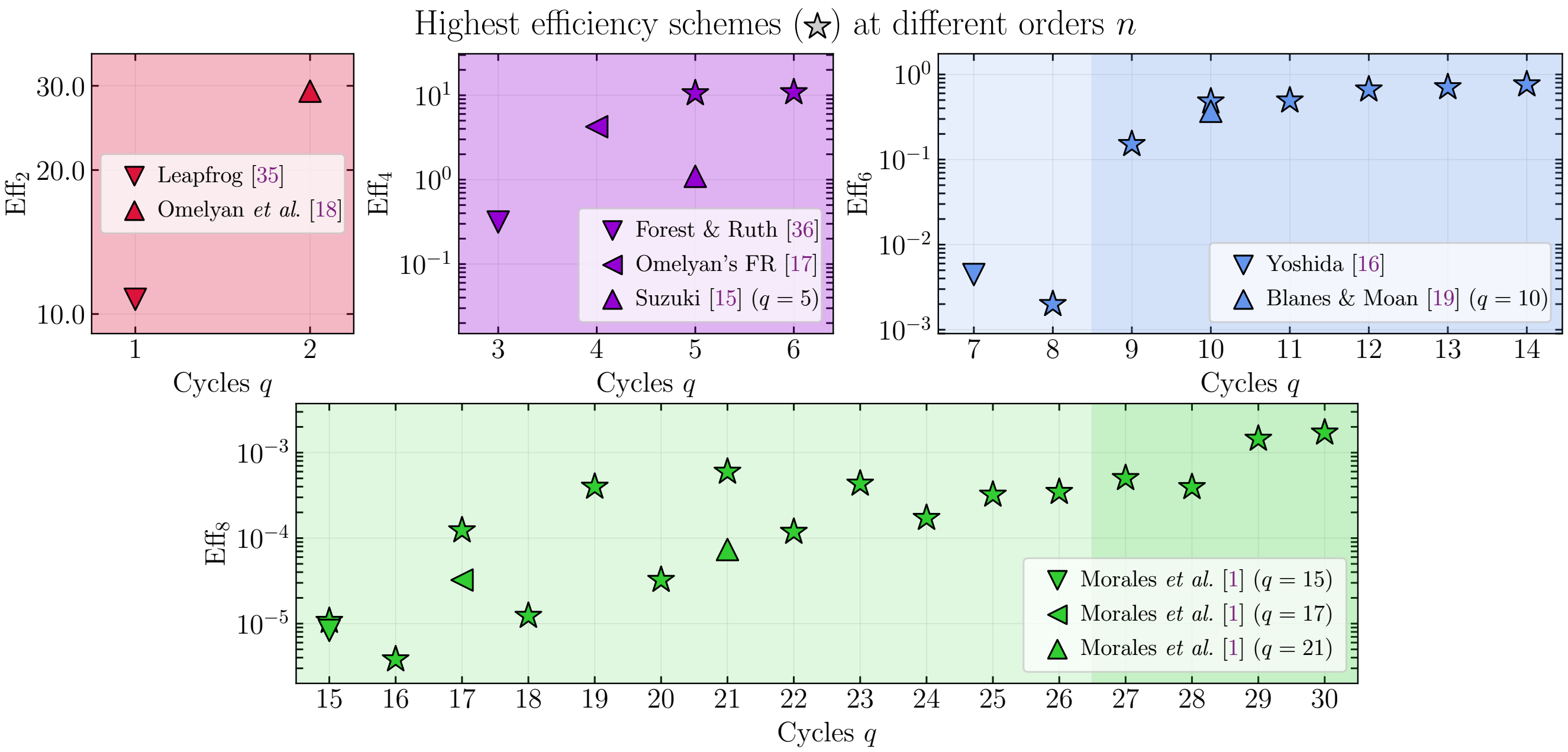}
	}
	\caption{Theoretical efficiency $\textrm{Eff}_{n}$~\eqref{eq:efficiency} of schemes from the literature (triangles) and our most efficient schemes (stars) for each number of cycles $q$ at orders $n \leq 8$, extending our previous results~\cite{Malezic_2026} to order $n = 8$.
         At orders $n = 6, 8$, the composition and generic regions are shaded in light and dark, respectively.
         In the composition region at $n = 8$, our schemes improve considerably on those by Morales \textit{et al}.~\cite{morales2022greatly}, especially at $q = 21$.
         At even $q$, the efficiency is mostly lower than at the neighbouring odd $q$, since the additional cycle adds no free parameter to the compositions.
         In the generic region, the efficiency rises further, and the most efficient schemes at $q = 29, 30$ are compositions.}\label{fig:theo_eff}
\end{figure*}

To compare schemes with different numbers of cycles $q$, we fix the computational cost, as in Ref.~\cite{Malezic_2026}.
Since each step consists of $q$ cycles, the cost scales as $q/h$, and we fix it by choosing the step size $h \propto q$.
With the global error scaling as $h^{n} \textrm{Err}_{n}$ (see Eq.~\eqref{eq:err_n8}), we define the efficiency

\begin{equation}
  \textrm{Eff}_{n} = \frac{1}{q^{n} \textrm{Err}_{n}},\label{eq:efficiency}
\end{equation}

\noindent
which only compares schemes of the same order $n$, since the error scales with a different power of $h$ for each order.

After the minimisation, we analyse the found minima and sort them by efficiency.
In \Cref{fig:theo_eff}, we compare our most efficient schemes with those from the literature, together with our schemes at orders $n \leq 6$ from Ref.~\cite{Malezic_2026}.
The stars mark our most efficient scheme at each number of cycles $q$, and the triangles the best known scheme from the literature.
At $6^{\textrm{th}}$ and $8^{\textrm{th}}$ order, the composition and generic regions are shaded in light and dark, respectively.
The new results of this work are shown in the bottom panel, where all previously known schemes are compositions with odd $q \leq 21$~\cite{YOSHIDA1990262, Kahan_Li_1997, Sofroniou_Spaletta_2005, morales2022greatly}, of which we show the most recent ones by Morales \textit{et al.}~\cite{morales2022greatly}.
Our framework extends this to even $q$, as well as beyond $q = 21$ up to $q = 30$, including generic schemes at $q \geq 27$, which cannot be constructed as compositions.

At $8^{\textrm{th}}$ order, our framework improves considerably on the known schemes, most notably at $q = 21$, where our best scheme is roughly $8$ times more efficient than~\cite{morales2022greatly}.
In the composition region, odd $q$ generally yield more efficient schemes than their even neighbours.
This is expected from \Cref{sec:manifold_structure}, since an even $q$ adds a cycle, and thus cost, without adding a free parameter to the compositions.
In the generic region, compositions still provide the most efficient schemes at $q = 27, 29$ and $30$, since their submanifold is of larger dimension than the generic manifold, and thus offers more freedom for the optimisation (see \Cref{tab:order8_regions}).
Only at $q = 28$, where the even $q$ penalises compositions, is a generic scheme the most efficient.
Since both manifolds grow in dimension with $q$, our search is likely incomplete at large $q$, and even more efficient schemes may exist there.
Theoretical efficiency, however, is not the only indicator of good performance, and we therefore perform numerical benchmarks before recommending scheme parameters.

\subsection{Numerical performance}\label{sec:numerical_performance}

\begin{figure*}[t]
	\centering
	\resizebox{\textwidth}{!}{%
		\input{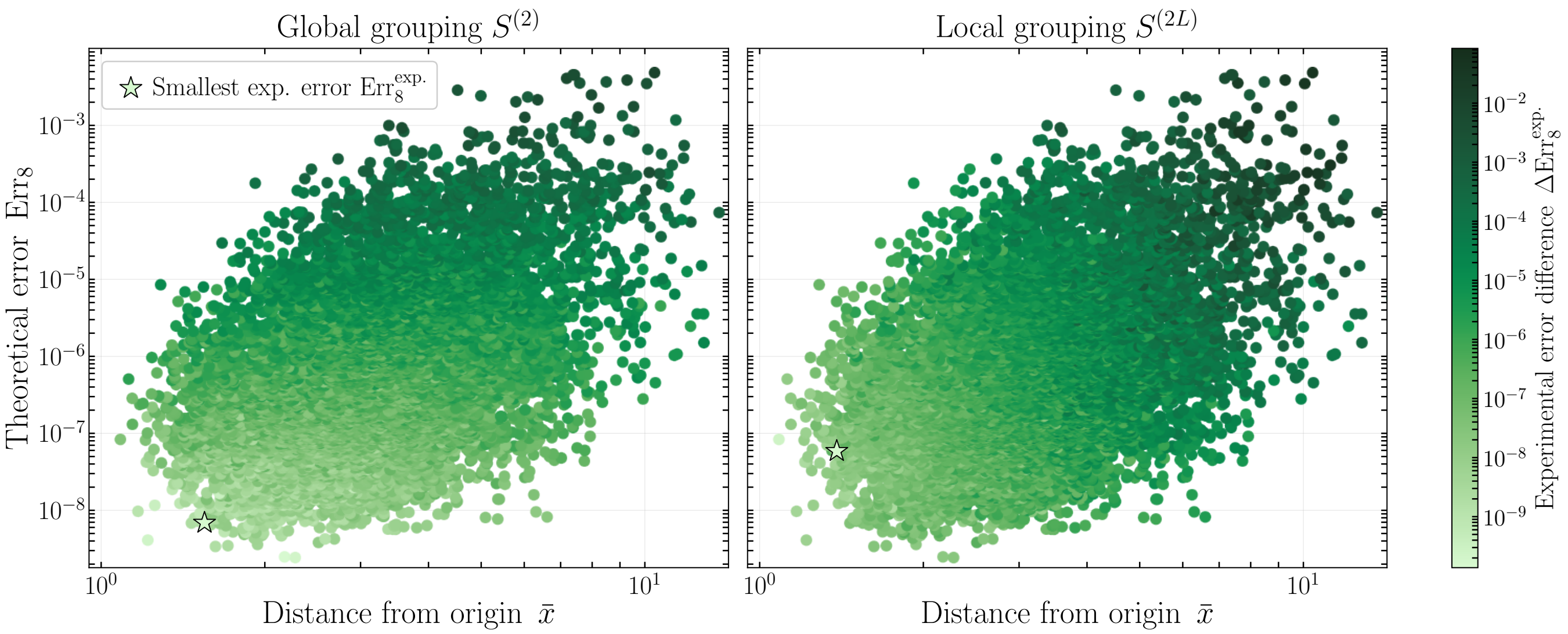}
	}
	\caption{Scheme comparison for generic decompositions at maximal number of cycles $q = 30$ in order $n = 8$.
           The schemes are classified according to the theoretical error $\textrm{Err}_{8}$~\eqref{eq:err_n8}, distance $\bar{x}$ from uniform scheme parameters~\eqref{eq:origin} and difference of the numerical error $\Delta \textrm{Err}_{8}^{\textrm{exp.}}$~\eqref{eq:diff_exp_err}.
           The latter is computed for a Heisenberg XZ spin chain $(L = 6)$ with open boundary conditions, and two product formula orderings time evolved until $t = 10$ at a fixed computational cost $\nicefrac{q}{h} = 10^{3}$.
           In the global one $S^{(2)}$ (left) it scales better with the theoretical error, while in the local grouping $S^{(2 L)}$ (right), practical performance correlates primarily with the distance from the origin $\bar{x}$.}\label{fig:comp_LG}
\end{figure*}

The theoretical efficiency measures the leading error of a scheme independently of the simulated system, and therefore does not map exactly onto the error observed in real simulations.
The error function $\textrm{Err}_{n}$ weights all nested commutators in the leading error term $O_{n+1}$ equally, whereas their actual contributions depend on the operators $A$ and $B$ of the simulated Hamiltonian.
Moreover, Hamiltonians of physical interest usually consist of more than two non-commuting parts, to which the schemes are applied through the generalisation of Ref.~\cite{Ostmeyer:2022}, introducing further commutators not captured by $\textrm{Err}_{n}$.
Finally, recent work shows that errors from successive steps can interfere, such that their accumulation over time also depends on the structure of the scheme~\cite{chen2024trottererrortimescaling, chen2024errorinterferencequantumsimulation}.
The model dependence cannot be removed in general, but we can study how the scheme parameters affect the error in practice, which is the focus of this section.

\flushbottom

In this work, we model a Heisenberg XZ spin chain with $L = 6$ sites and open boundary conditions, which has the following Hamiltonian

\begin{equation}
  H = \sum_{i = 1}^{L-1} J \left( \sigma_{i}^{x} \sigma_{i+1}^{x} + \sigma_{i}^{z} \sigma_{i+1}^{z} \right) + \sum_{i=1}^{L} h_{i} \sigma_{i}^{z},\label{eq:XZ_hamiltonian}
\end{equation}

\noindent
where we denote the Pauli matrices at site $i$ as $\sigma_{i}^{\alpha}$, with the coupling set to unity $J = 1$.
To break the symmetries of the model, we add a random magnetic field in the $z$-direction, with $h_{i}$ drawn uniformly from $[-0.1, 0.1]$ and kept fixed for all simulations.

\begin{figure*}[t]
	\centering
	\resizebox{0.94\textwidth}{!}{%
		\input{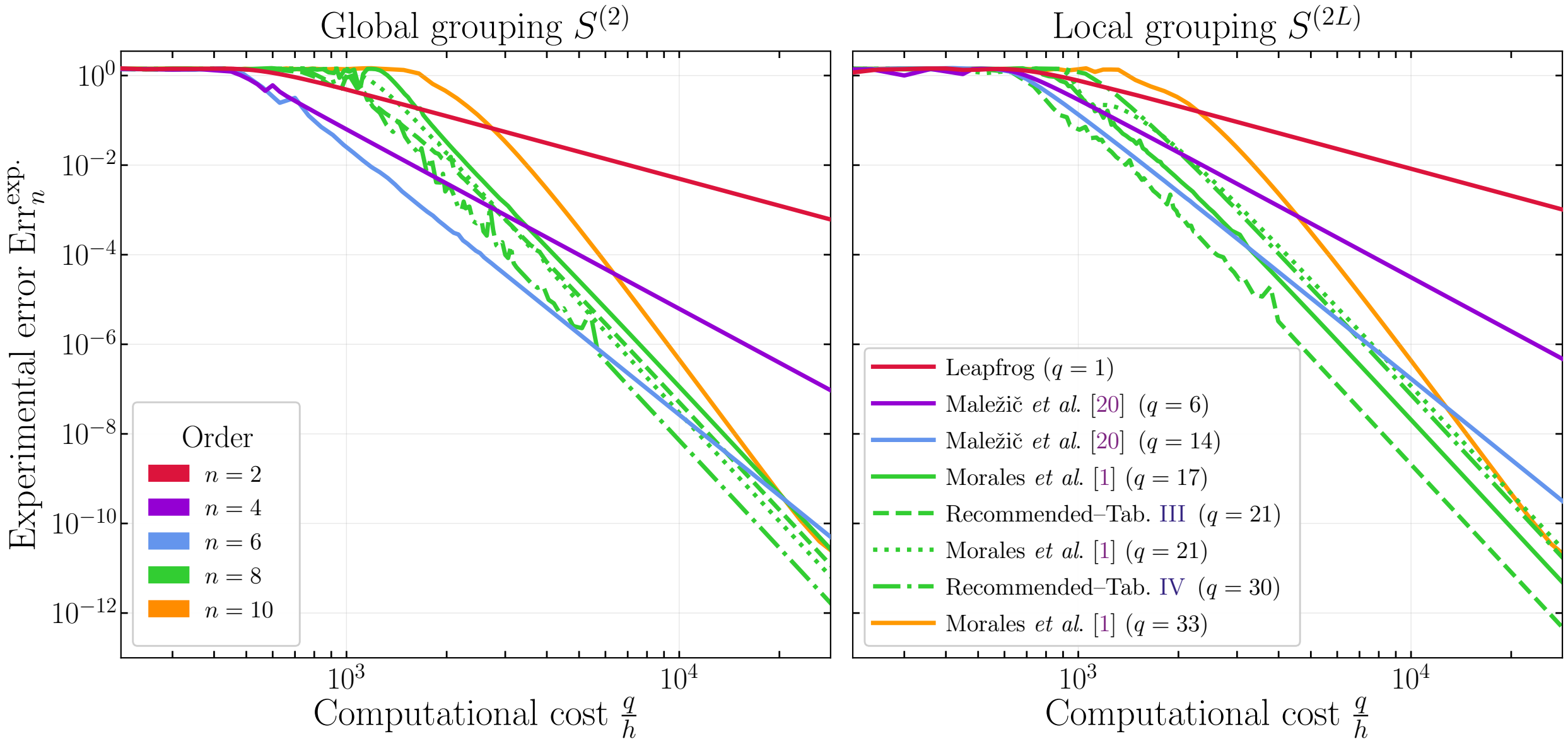}
	}
	\caption{Experimental error $\textrm{Err}_{n}^{\textrm{exp.}}$~\eqref{eq:exp_error} of a Heisenberg XZ spin chain with open boundary conditions and length $L = 6$, plotted against the computational cost $\nicefrac{q}{h}$ for the global ordering $S^{(2)}$ (left) and the local ordering $S^{(2L)}$ (right).
           With a focus on long-time evolution, we set the total time $t = 100$ and vary the time step $h$.
           We compare the Leapfrog scheme, our recommended schemes at $n = 4, 6$~\cite{Malezic_2026} and our novel schemes at $n = 8$ (\Cref{tab:scheme_q21,tab:scheme_q30}) to those by Morales \textit{et al}.~\cite{morales2022greatly} at $n = 8, 10$.
           Each of our two schemes reaches a lower error in one ordering, and remains comparable in the other.}\label{fig:best_cost}
\end{figure*}

With many local (non-)commuting operators it is possible to split the Hamiltonian in many different ways~\cite{liu2026orderingawaretheorytrottererror}.
Since each term acts on at most two neighbouring sites, it can be implemented directly as a local gate.
A natural grouping follows the chain bond by bond $(i, i+1)$,

\begin{equation}
  \begin{aligned}
    S_{n}^{(2 L)} (h) &= e^{-i c_{1} h H_{1}^{x}} e^{-i c_{1} h H_{1}^{z}} e^{-i c_{1} h H_{2}^{x}} e^{-i c_{1} h H_{2}^{z}} \cdots\\
                  &\times e^{-i d_{q} h H_{2}^{z}} e^{-i d_{q} h H_{2}^{x}} e^{-i d_{q} h H_{1}^{z}} e^{-i d_{q} h H_{1}^{x}},
  \end{aligned}\label{eq:local_PF}
\end{equation}

\noindent
where $H_{i}^{\alpha} = J \sigma_{i}^{\alpha} \sigma_{i+1}^{\alpha} + \delta_{\alpha z} h_{i} \sigma_{i}^{\alpha}$ denote the local terms, and the product formula is written in the ramp notation, which generalises naturally to $\Lambda$ stages~\cite{Ostmeyer:2022}.
Since this product formula consists of $\Lambda = 2L-1$ stages, we refer to it as the local ordering $S^{(2L)}$.
A different approach exploits commutativity: since all $H_{i}^{x}$ commute with each other, as do all $H_{i}^{z}$, they can be grouped into only $\Lambda = 2$ stages,

\begin{equation}
  \begin{aligned}
    S_{n}^{(2)} (h) &= \left( \prod_{i = 1}^{L} e^{-i c_{1} h H_{i}^{x}} \right) \left( \prod_{i = 1}^{L} e^{-i c_{1} h H_{i}^{z}} \right) \cdots\\
                &\times \left( \prod_{i = L}^{1} e^{-i d_{q} h H_{i}^{z}} \right) \left( \prod_{i = L}^{1} e^{-i d_{q} h H_{i}^{x}} \right),
  \end{aligned}\label{eq:global_PF}
\end{equation}

\noindent
with each stage spanning the whole chain, and we therefore refer to it as the global ordering $S^{(2)}$.

We compare our $8^{\textrm{th}}$ order schemes to the exact time evolution $U(t) = \exp(- i t H)$.
Accordingly, the Trotterised evolution $\tilde{U}(t) = \big[ S_{n}^{(\Lambda)}(h) \big]^{t/h}$ consists of $t/h$ steps of the product formula.
We define the experimental error through the normalised Frobenius norm of $U - \tilde{U}$, which we compute as

\begin{equation}
  \textrm{Err}_{n}^{\textrm{exp.}} = \sqrt{\frac{1}{N} \sum\nolimits_{v} \big\lVert (U - \tilde{U}) |v\rangle \big\rVert^{2}},\label{eq:exp_error}
\end{equation}

\noindent
where the sum runs over the $N = 2^{L}$ computational basis states $|v\rangle$, and $\lVert \cdot \rVert$ denotes the Euclidean norm.

Our first numerical experiment is to simulate the new product formulas at a fixed computational cost $\nicefrac{q}{h} = 10^{3}$ for a fixed total time $t = 10$.
By classifying schemes according to their theoretical error $\textrm{Err}_{8}$~\eqref{eq:err_n8} and the distance from the origin $\bar{x}$~\eqref{eq:origin}, as in \Cref{fig:manifolds_q17-18}.
To these theoretical classifiers however, we can add the experimental error~\eqref{eq:exp_error} as another dimension.
This allows us to understand which of the two quantities impacts the efficiency in practice.

\raggedbottom

In \Cref{fig:comp_LG} we present such a result for the maximal number of cycles $q = 30$ in both the local $S^{(2 L)}$ and global $S^{(2)}$ product formula ordering (or grouping).
Since the two orderings reach different absolute errors, we colour each scheme by its experimental error relative to the smallest one, $\textrm{Err}_{8}^{\star}$, among all schemes in the same ordering,

\begin{equation}
  \Delta \textrm{Err}_{8}^{\textrm{exp.}} = \textrm{Err}_{8}^{\textrm{exp.}} - \textrm{Err}_{8}^{\star},\label{eq:diff_exp_err}
\end{equation}

\noindent
which places both orderings on a common colour scale.
In the global ordering $S^{(2)}$, the experimental error correlates mainly with $\textrm{Err}_{8}$, with the best schemes at the bottom.
The experimental error then grows in the direction of a larger $\textrm{Err}_{8}$ and slightly towards a larger distance $\bar{x}$.
This is expected, since $\textrm{Err}_{8}$ is derived for exactly two operators, and therefore captures all error terms of a product formula with $\Lambda = 2$ stages.
In the local ordering $S^{(2L)}$, on the other hand, the experimental error correlates mainly with the distance $\bar{x}$, with the best schemes on the left, close to uniform scheme parameters.
With $\Lambda = 2L-1$ stages, many commutators between local terms are not captured by $\textrm{Err}_{8}$, and their contribution presumably grows with the size of the scheme parameters, which favours nearly uniform schemes.
This is in line with recent findings that the structure of a scheme affects the error accumulation~\cite{chen2024trottererrortimescaling, chen2024errorinterferencequantumsimulation}.
Whether uniformity generally plays a larger role for product formulas with many stages requires further study, but we take it into account for our recommendations.

For the next simulation, we increase the total time to $t = 100$ to study long-time error accumulation, and vary the computational cost $\nicefrac{q}{h}$, which yields the scaling plots shown in \Cref{fig:best_cost}.
As reference, we choose the Leapfrog scheme, our recommended schemes from Ref.~\cite{Malezic_2026} at $n = 4, 6$, and the best schemes by Morales \textit{et al}.~\cite{morales2022greatly} at $n = 8, 10$.
Out of our novel $8^{\textrm{th}}$ order schemes, we select a highly efficient scheme from the composition region at $q = 21$, and a generic scheme at the maximal number of cycles $q = 30$.
Both combine a high theoretical efficiency $\textrm{Eff}_{8}$ with a comparatively small distance $\bar{x}$, and their parameters are given in Tables~\ref{tab:scheme_q21} and~\ref{tab:scheme_q30}, respectively, while we discuss their choice in \Cref{sec:scheme_recommendations}.
We do not select the most efficient compositions at $q = 29, 30$, since, despite an efficiency roughly three times higher, their large distances $\bar{x} \geq 2$ lead to larger errors in both orderings.

For the scaling analysis shown in \Cref{fig:best_cost} we again investigate the two groupings $S^{(2)}$ and $S^{(2 L)}$.
Our two selected schemes perform well in both, but each excels in a different one.
The highly efficient and uniform scheme at $q = 21$ cycles (see \Cref{tab:scheme_q21}) performs best in the local $S^{(2 L)}$ grouping, where it reaches a lower error than the formulas by Morales \textit{et al}.~\cite{morales2022greatly}.
In this ordering, it also outperforms our recommended schemes at lower orders $n = 4, 6$~\cite{Malezic_2026} across the computational cost in the scaling region.
In the global ordering $S^{(2)}$, the $q = 21$ scheme is only comparable to those by Morales \textit{et al}., while our second scheme at the maximal number of cycles $q = 30$ (see \Cref{tab:scheme_q30}) outperforms them slightly.

Comparing both orderings, the $q = 21$ scheme in the local ordering $S^{(2L)}$ performs best overall, closely followed by the generic scheme at $q = 30$ in the global ordering $S^{(2)}$.
We attribute this to its high theoretical efficiency and small distance $\bar{x}$.
At lower orders, on the other hand, our recommended schemes perform better in the global ordering than in the local one.
We attribute this to their choice as a compromise between efficiency and uniformity, and a stronger preference for uniformity may be beneficial in local orderings.
This is in line with the observation that, for high-order product formulas, the best choice of scheme parameters can depend on the ordering~\cite{liu2026orderingawaretheorytrottererror}.
Based on these numerical experiments, we discuss our recommendations for $8^{\textrm{th}}$ order schemes in the following subsection.

\subsection{Scheme recommendations}\label{sec:scheme_recommendations}

Having optimised the scheme parameters and analysed their error accumulation in time evolution simulations, we can now give $8^{\textrm{th}}$ order scheme recommendations, based on both theoretical and practical criteria.
On the one hand, we look for schemes with a high efficiency $\textrm{Eff}_{8}$~\eqref{eq:efficiency}, i.e.\ a small leading error at fixed computational cost.
On the other hand, we prefer uniform schemes with a small distance $\bar{x}$, which showed a favourable error accumulation in our simulations, especially in the local ordering.
Based on these criteria, we have selected a few schemes at each number of cycles, which are available in our repository~\cite{markomalezic_2026_18347430}.
Since practical performance depends on the model and the ordering (see \Cref{sec:numerical_performance}), we additionally favour robust schemes, which never perform badly and ideally are favourable in some settings, as observed for our two orderings.
As this cannot be guaranteed for all systems, our general recommendation relies largely on the theoretical criteria above.

At $8^{\textrm{th}}$ order, we recommend two schemes.
The first is a highly efficient scheme from the composition region at $q = 21$, whose parameters are given in \Cref{tab:scheme_q21}.
Besides its high efficiency, it is also one of the most uniform schemes we found.
Its mirror symmetry $c_{j} = c_{q+1-j}$ identifies it as a composition of Leapfrog steps, the only type of scheme in this region (see Sec.~\ref{sec:manifold_structure}).
These properties likely make it a good choice for many different models and operator orderings.
In the Heisenberg XZ chain (see Fig.~\ref{fig:best_cost}), it outperforms the schemes by Morales \textit{et al}.~\cite{morales2022greatly} in the local ordering, and is comparable to them in the global one.

\begin{table}[h!]
  \centering
	\caption{Recommended scheme parameters $c_{j}$ for a highly efficient scheme at $q = 21$ and $n = 8$ with one of the smallest distances $\bar{x}$.
           Its mirror symmetry $c_{j} = c_{q+1-j}$ identifies it as a Yoshida-type composition scheme~\eqref{eq:composition}.
           Due to its efficiency and uniformity, it is a good candidate for a wide range of models and orderings.}\label{tab:scheme_q21}
  \begin{tabular}{C{0.5cm}S[table-format=-1.18, group-digits=none, table-column-width=5.75cm]}
		\toprule
		\multicolumn{2}{c}{$\textrm{Eff}_{8} = \num[round-mode=figures,round-precision=4]{5.965037e-4}$} \\
		\multicolumn{2}{c}{$\bar{x} = \num[round-mode=figures,round-precision=3]{0.689251}$} \\
		\midrule
		$j$ &                   {$c_{j} = d_{q+1-j}$}   \\
		\midrule
		1  &  0.041644811759980006 \\
		2  &  0.082401316690625129 \\
		3  & -0.100053679877075558 \\
		4  &  0.056276640904364281 \\
		5  &  0.054789293427501641 \\
		6  &  0.052417975744997715 \\
		7  &  0.202143137014514557 \\
		8  & -0.156387050011859596 \\
		9  &  0.047476878333309942 \\
		10 &  0.077958103121855915 \\
		11 & -0.217334854216428036 \\
		12 &  0.077958103121855915 \\
		13 &  0.047476878333309942 \\
		14 & -0.156387050011859596 \\
    15 &  0.202143137014514557 \\
    16 &  0.052417975744997715 \\
    17 &  0.054789293427501641 \\
    18 &  0.056276640904364281 \\
    19 & -0.100053679877075558 \\
    20 &  0.082401316690625129 \\
    21 &  0.041644811759980006 \\
		\bottomrule
	\end{tabular}
\end{table}

The second recommended scheme is a highly efficient generic scheme at the maximal number of cycles $q = 30$, whose parameters are given in \Cref{tab:scheme_q30}.
As a generic scheme, it does not exhibit the mirror symmetry of the first recommendation.
Although generic schemes are not restricted to compositions, the most efficient generic scheme we found at $q = 30$ is only slightly more efficient than the one in \Cref{tab:scheme_q21}, and has a comparatively large distance $\bar{x}$.
We therefore recommend a more uniform scheme at $q = 30$, which is still less efficient and less uniform than our first recommendation, but performs well in our benchmarks, especially in the global ordering.
In general, uniform schemes were harder to find in the generic region.
Our minimiser also does not reach the compositions that still exist there (see \Cref{sec:manifold_structure}), and a search restricted to Yoshida-type schemes could reveal more efficient and uniform schemes.

\begin{table}[h!]
  \centering
	\caption{Recommended scheme parameters $c_{j}$ for a generic scheme at the maximal number of cycles $q = 30$ at order $n = 8$.
         Although it is less efficient and uniform than our recommendation at $q = 21$, it performs well in our benchmarks, especially in the global ordering, where uniformity has less impact.}\label{tab:scheme_q30}
  \begin{tabular}{C{0.5cm}S[table-format=-1.18, group-digits=none, table-column-width=5.75cm]}
		\toprule
		\multicolumn{2}{c}{$\textrm{Eff}_{8} = \num[round-mode=figures,round-precision=4]{3.685429e-4}$} \\
		\multicolumn{2}{c}{$\bar{x} = \num[round-mode=figures,round-precision=4]{1.220032}$} \\
		\midrule
		$j$ & {$c_{j} = d_{q+1-j}$}\\
		\midrule
		1  &  0.026395999625833815 \\
		2  & -0.233684379758421046 \\
		3  & -0.347051470763567071 \\
		4  & -0.274493109455153539 \\
		5  &  0.061094592879806627 \\
		6  &  0.080452813774411269 \\
		7  & -0.063568811420699228 \\
		8  &  0.079026700770564284 \\
		9  &  0.059258569653719090 \\
		10 &  0.101028778330863239 \\
		11 & -0.164724202482292914 \\
		12 & -0.153554503905898948 \\
		13 &  0.107485850556923307 \\
		14 & -0.132867087173779791 \\
		15 & -0.074760981570922924 \\
		16 &  0.153609834123292238 \\
		17 &  0.024200829485005276 \\
		18 & -0.155771661397294425 \\
		19 &  0.089897001564395437 \\
		20 &  0.185309492367605244 \\
		21 &  0.093200666104659380 \\
		22 &  0.092333475195637504 \\
		23 &  0.050724396546212469 \\
		24 &  0.057234606768893276 \\
		25 & -0.114753172414141125 \\
		26 &  0.033620005282797918 \\
		27 &  0.217672800411770034 \\
		28 &  0.344599868646588370 \\
		29 &  0.286949675877411436 \\
		30 &  0.071133422375779604 \\
		\bottomrule
	\end{tabular}
\end{table}

Besides the two presented schemes, we find further competitive schemes at order $n = 8$, some favouring efficiency and others uniformity, which may perform better in certain scenarios, and we provide them in \Cref{app:additional_schemes}.
Motivated by the role of uniformity in the local ordering, we also revisit orders $n = 4, 6$, where more uniform schemes improved on our previous recommendations in some of our tests, and we include them in the same appendix.
All found schemes at order $n = 8$, along with a few recommendations at each number of cycles $q$ and their properties, are archived in our repository~\cite{markomalezic_2026_18347430}, most of them minimised to quadruple precision (36 significant digits).
The repository also contains our program, updated with the new methodology and quadruple-precision support.


\section{Quantum circuits for time evolution}\label{sec:quantum_circuits}

Although product formulas are used across many fields, they have recently come into the spotlight for applications in quantum computing.
Quantum computers promise to simulate the time evolution of large quantum systems with resources that scale only polynomially in system size~\cite{Feynman1982, Lloyd1996}.
Since each factor $e^{i c_{j} A_{k}}$ of a product formula can be implemented directly as a short sequence of quantum gates, without ancillas or block encodings, product formulas are natural candidates for this task.
For long-time dynamics, however, the number of Trotter steps, and thus the gate count, grows quickly, which makes efficient high-order schemes particularly attractive.
In this section, we therefore examine the applicability of our novel schemes to quantum circuits.

Our benchmarks in \Cref{sec:numerical_performance} were performed classically, without noise in any stage $e^{i c_{j} A_{k}}$.
They represent the ideal case of a fault-tolerant quantum computer, and give hope that efficient high-order product formulas could be useful once fault tolerance is achieved.
Until then, we must deal with decoherence and noisy gates, which we focus on in the following.

\begin{figure*}[t]
	\centering
	\resizebox{0.955\textwidth}{!}{%
		\input{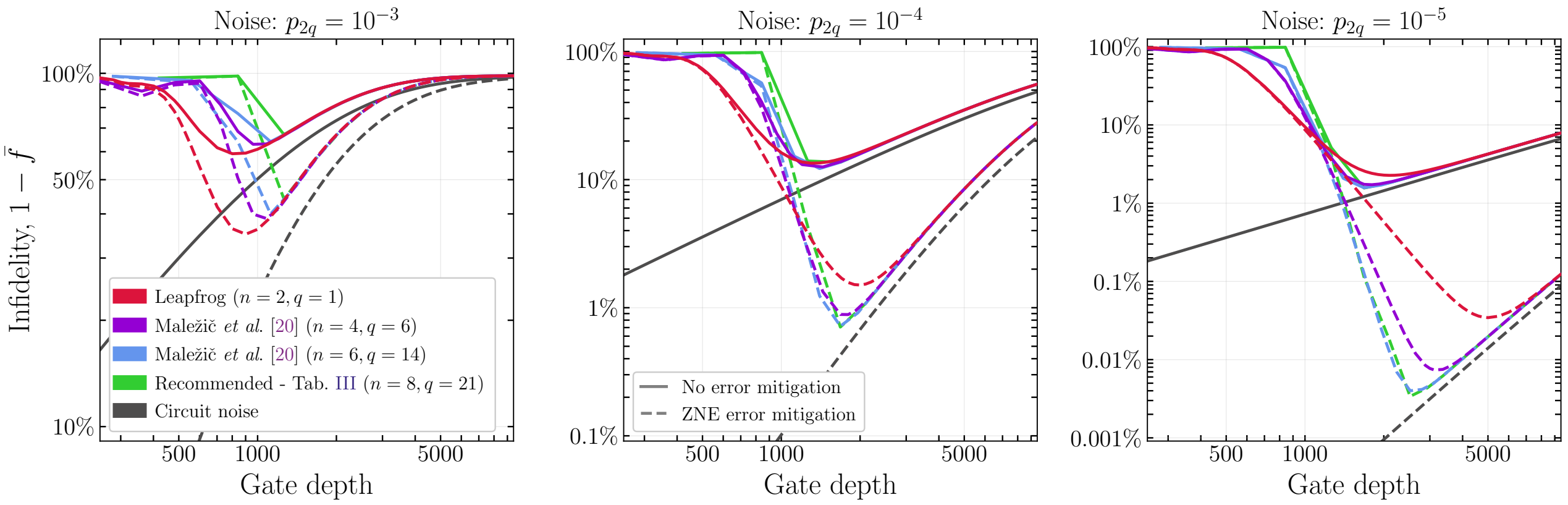}
	}
	\caption{Infidelity $1 - \bar{f}$~\eqref{eq:avg_fid} against gate depth for the Heisenberg XZ chain ($L = 6$, open boundaries, $t = 10$) in the local ordering $S^{(2L)}$, at three 2-qubit depolarising noise levels $p_{2q}$ with $p_{1q} = p_{2q}/10$.
           The gate depth is varied through the number of Trotter steps.
           Solid lines show the results without and dashed lines with ZNE error mitigation, and the grey lines show the circuit noise of the zero-angle Leapfrog circuits.
           At the state-of-the-art NISQ~\cite{Yin2026} noise level $p_{2q} = 10^{-3}$ (left), Leapfrog remains the best choice.
           At $p_{2q} = 10^{-4}$ (middle), all schemes perform similarly well without mitigation, while ZNE reveals a two-fold improvement of the higher orders.
           At $p_{2q} = 10^{-5}$ (right), the advantage grows to roughly an order of magnitude (see \Cref{tab:noisy_minima}).}\label{fig:noisy_grid}
\end{figure*}

To simulate noisy circuits, we use Qiskit~\cite{qiskit2024} together with \texttt{qiskit-omelyan}~\cite{qiskit-omelyan}, a plugin we developed to construct product formulas based on the Omelyan method.
Whereas core Qiskit only provides the recursive Suzuki construction for higher orders, whose number of stages grows exponentially with the order, our plugin constructs product formulas with user-defined scheme parameters at any number of cycles $q$.
It includes several established schemes from the literature, together with our recommended schemes for orders $n = 4, 6, 8$, and is implemented as a high-level synthesis method for Pauli evolution gates, which integrates directly into the standard Qiskit transpilation workflow.

Once the circuit is built, we simulate it with the density-matrix method of Qiskit Aer, applying noise to every Pauli rotation gate.
Rather than a hardware-specific noise model, we use a simple depolarising channel after every gate, with error probabilities $p_{1q}$ and $p_{2q}$ for 1- and 2-qubit gates, respectively.
Since 2-qubit gates are generally harder to implement and noisier, we set $p_{1q} = p_{2q}/10$.

For our noisy benchmark, we again simulate the Heisenberg XZ spin chain ($L = 6$) with open boundaries.
For clarity, we choose a total time $t = 10$, and restrict ourselves to the local ordering $S^{(2 L)}$.
Instead of the experimental error defined in Eq.~\eqref{eq:exp_error}, we compute the fidelity between exactly time-evolved and Trotterised (noisy) states.
Since the exactly evolved states $|v(t)\rangle = e^{-i H t} |v\rangle$ are pure, the (squared) Uhlmann fidelity reduces to

\begin{equation}
  f_{v} = \langle v(t) | \rho_{v} | v(t) \rangle,\label{eq:single_fidelity}
\end{equation}

\noindent
where $\rho_{v}$ represents the density matrix obtained by evolving $|v\rangle$ through the Trotterised, noisy circuit.
For a better estimate, we again average over all $N = 2^{L}$ computational basis states $|v\rangle$,

\begin{equation}
  \bar{f} = \frac{1}{N} \sum_{v} f_{v}.\label{eq:avg_fid}
\end{equation}

Before discussing the results of our noisy simulations, we introduce the error-mitigation technique we employ, namely zero-noise extrapolation (ZNE)~\cite{PhysRevLett.119.180509, PhysRevX.7.021050}.
We amplify the noise by local unitary folding~\cite{GiurgicaTiron2020}, replacing every gate $G$ of the transpiled circuit by $G (G^{\dagger} G)^{k}$ with $k = 0, 1, 2$, which yields noise scale factors $\lambda = 1, 3, 5$.
For each initial state $|v\rangle$ we then extrapolate $f_{v}(\lambda)$ to $\lambda = 0$ by Richardson extrapolation, i.e.\ a quadratic polynomial through the three points, as implemented in Mitiq~\cite{LaRose2022mitiqsoftware}, which we have chosen since it assumes no particular noise model.
Note that ZNE targets the noiseless Trotterised circuit, hence it cannot remove the Trotter error.
Moreover, the polynomial extrapolation becomes unreliable for deep circuits, where the fidelity decays approximately exponentially with the number of gates and the folded circuits approach the maximally mixed state.
We stress that our ZNE setting is idealised: each folded gate carries the same depolarising noise as the original, so folding amplifies the noise by exactly $\lambda$, and our density-matrix simulation is free of shot noise, which the extrapolation would otherwise amplify.
Our ZNE results should therefore be understood as an optimistic estimate of what mitigation could achieve on hardware.
The scripts to construct such time evolution circuits for the Heisenberg model with our plugin~\cite{qiskit-omelyan} can be found in our repository~\cite{markomalezic_2026_18347430}.

We present the results of our noisy simulations in \Cref{fig:noisy_grid} for three values of the 2-qubit depolarising noise $p_{2q}$.
To allow a direct comparison with the noiseless scaling in \Cref{fig:best_cost}, we plot the infidelity $1 - \bar{f}$ against the gate depth, which we vary through the number of Trotter steps.
We compare the Leapfrog scheme ($n = 2$) with our recommended schemes for $n = 4$ and $n = 6$ from Ref.~\cite{Malezic_2026}, and the new $n = 8$ scheme from \Cref{tab:scheme_q21}.
We also plot the circuit noise, obtained by simulating the Leapfrog circuits with all rotation angles set to zero.
Since these circuits ideally implement the identity on computational basis states, they serve as a reference for the noise accumulated at a given depth, independent of the Trotter error.
However, they slightly underestimate the noise, since $Z$-type errors leave computational basis states invariant.
In addition to the noisy scaling plots, we tabulate the minimal infidelity values for each curve in \Cref{tab:noisy_minima}.

The three different noise levels in \Cref{fig:noisy_grid} represent a potential timeline for the applicability of Trotter--Suzuki schemes in quantum computation.
In the current \textit{noisy intermediate-scale quantum} (NISQ) era~\cite{Preskill2018quantumcomputingin,Yin2026}, where even the best devices are limited to $p_{2q} \gtrsim 10^{-3}$, accurate long-time evolution remains unattainable due to large gate depths.
In this regime, the Leapfrog scheme stays the best option among Trotterizations due to its simplicity both with and without error mitigation, although the infidelity remains high.
Alternative methods, such as constant-depth multi-product formulas (MPFs)~\cite{bark2026constantdepthmultiproductformulatrotter} or (partially) randomised algorithms~\cite{granet2023continuous,Hayata:2026qcz,Gunther:2025txp,Murota:2026cyh}, address this limitation.
An extensive list of state-of-the-art approaches to quantum simulations on NISQ devices can be found in the review~\cite{Ostmeyer:2026ckx}.

As the hardware improves, the application of higher-order decompositions becomes more viable.
At $p_{2q} = 10^{-4}$ without error mitigation, all schemes perform roughly the same, but the error is too large for reliable results.
With error mitigation, the noise is pushed down, revealing the advantage of higher orders $n \geq 4$, which halve the error compared to Leapfrog (see Tab.~\ref{tab:noisy_minima}).
At $p_{2q} = 10^{-5}$, the higher-order schemes improve on Leapfrog already without error mitigation, and with ZNE, the differences become pronounced.
The best result is obtained by our novel $n = 8$ scheme, closely followed by the recommended $n = 6$ scheme, both reaching an infidelity roughly an order of magnitude below the Leapfrog minimum, at half the gate depth.
The recommended $n = 4$ scheme lies in between, with a roughly 5 times lower infidelity than Leapfrog.
The higher-order schemes are therefore both more accurate (see Tab.~\ref{tab:noisy_minima}) and more cost-efficient (see Fig.~\ref{fig:noisy_grid}).
Although the figure presents an idealised scenario, it suggests that already in the era of early fault-tolerant quantum computing, high-order product formulas could provide a simple path towards the efficient simulation of long-time dynamics.

\begin{table}[t]
  \centering
  \caption{Minimal infidelity $1 - \bar{f}$ (in \%) for each scheme at order $n$ and 2-qubit depolarising noise $p_{2q}$ in \Cref{fig:noisy_grid}, without and with ZNE error mitigation.
           As the noise level $p_{2q}$ decreases, higher-order product formulas become more advantageous, making them viable for early fault-tolerant simulation.}\label{tab:noisy_minima}
  \begin{tabular*}{\columnwidth}{@{\extracolsep{\fill}}cccccc}
    \toprule
    $p_{2q}$ & ZNE & $n = 2$ & $n = 4$ & $n = 6$ & $n = 8$ \\
    \midrule
    \multirow{2}{*}{$10^{-3}$} & no & $\mathbf{59.2}$ & $63.0$ & $64.1$ & $67.1$ \\
                               & yes & $\mathbf{35.0}$ & $38.8$ & $40.0$ & $44.1$ \\
    \midrule
    \multirow{2}{*}{$10^{-4}$} & no & $13.4$ & $12.6$ & $\mathbf{12.2}$ & $13.7$ \\
                               & yes & $1.50$ & $0.88$ & $0.72$ & $\mathbf{0.71}$ \\
    \midrule
    \multirow{2}{*}{$10^{-5}$} & no & $2.27$ & $1.72$ & $1.57$ & $\mathbf{1.56}$ \\
                               & yes & $0.034$ & $0.0074$ & $0.0040$ & $\mathbf{0.0034}$ \\
    \bottomrule
  \end{tabular*}
\end{table}


\section{Conclusion}\label{sec:conclusion}

In this work, we have derived new highly efficient order $n = 8$ Suzuki-Trotter decompositions, see \Cref{tab:scheme_q21,tab:scheme_q30}.
Higher orders become particularly viable for long time evolutions since the cost to reach a given error $\epsilon$ scales as $\order{t^{1+\frac1n}\epsilon^{-\frac1n}}$ with the total time $t$.
In the process of deriving and benchmarking the new schemes, we have discovered that it depends on the Hamiltonian whether a small leading order error or maximally uniform scheme parameters lead to optimal performance.

We have extended our framework for the ab initio construction of high-order product formulas from Ref.~\cite{Malezic_2026} to order $n = 8$ in \Cref{sec:term_construction}.
We identified the $8^{\textrm{th}}$ order basis commutators $F_{k}$~\eqref{eq:basis_n8} and derived their coefficients $\epsilon_{k}$~\eqref{eq:eps56} through recursive formulae, which allow us to minimise the $O_{9}$~\eqref{eq:error_term_O9} error term.
By examining the constraint Jacobian in \Cref{sec:manifold_structure}, we confirmed the manifold structure of order $n = 8$ solutions, for which only composition schemes had been known so far.
While our framework was designed to find generic schemes, it could still identify Yoshida-type schemes in the composition region.
As we approach the generic region, however, the Jacobian of the general framework becomes progressively ill-conditioned, and the number of found minima declines (see \Cref{fig:count_condition}).
By imposing the mirror symmetry of compositions directly on the scheme parameters, we restored a low condition number, and found more minima at every number of cycles beyond $q = 16$ (see Sec.~\ref{sec:optim_improve}).

With these improvements, our approach established novel schemes with improved efficiencies $\textrm{Eff}_{8}$~\eqref{eq:efficiency} in the composition region (see Fig.~\ref{fig:theo_eff}).
Furthermore, we found generic schemes at $q \geq 27$, which had not been studied until now, and whose efficiencies are comparable to the best compositions at $q = 21$.
Since theoretical efficiency does not map exactly onto practical performance, we benchmarked our new schemes on the Heisenberg XZ spin chain of length $L = 6$ in \Cref{sec:numerical_performance}.
As part of our numerical experiments, we compared two operator orderings, the local $S^{(2L)}$~\eqref{eq:local_PF} and global $S^{(2)}$~\eqref{eq:global_PF} groupings.
In the global ordering, we observed that the experimental error correlates with the efficiency $\textrm{Eff}_{n}$, since its $\Lambda = 2$ stages are fully captured by $\textrm{Err}_{n}$, while in the local ordering it correlates mainly with the distance from the origin $\bar{x}$~\eqref{eq:origin} (see Fig.~\ref{fig:comp_LG}).
To compare our schemes to those by Morales \textit{et al}.~\cite{morales2022greatly}, we studied the scaling of the accumulated error $\textrm{Err}_{n}^{\textrm{exp.}}$ with the computational cost $\nicefrac{q}{h}$ in \Cref{fig:best_cost}.
Our recommended composition scheme at $q = 21$ performed best in the local grouping $S^{(2L)}$, while our generic scheme at $q = 30$ performed best in the global one $S^{(2)}$; their parameters are given in Tables~\ref{tab:scheme_q21} and~\ref{tab:scheme_q30}.
We also discussed additional schemes in \Cref{app:additional_schemes}, and gave our recommendations for every number of cycles $q$ in our repository~\cite{markomalezic_2026_18347430}.

Finally, in \Cref{sec:quantum_circuits}, we focused on the applicability of Trotter-Suzuki schemes in quantum circuits.
We simulated the Heisenberg XZ chain as before, but with noise applied to each gate that implements a Trotter step.
In order to do this efficiently, we developed a plugin for Qiskit~\cite{qiskit2024}, which we named \texttt{qiskit-omelyan}~\cite{qiskit-omelyan} after the approach of Omelyan \textit{et al}.~\cite{OMELYAN2003272}.
With the plugin, we synthesised evolution circuits and applied a simple noise model with a depolarising level $p_{2q}$ for 2-qubit gates.
We also studied the effect of error mitigation using zero-noise extrapolation (ZNE).
For three noise levels $p_{2q}$, we calculated the infidelity of our evolved states and plotted it against the gate depth required (see Fig.~\ref{fig:noisy_grid}).
We found that at noise levels of current state-of-the-art hardware, the Leapfrog scheme remains the best choice.
However, at lower noise levels our recommended schemes become competitive, which makes them applicable to long-time dynamics as hardware improves.

The adaptation of our framework to composition schemes will become essential, if we are to extend it to order $n = 10$, where generic schemes only appear at $q \geq 83$.
The main bottleneck there, however, lies in the derivation of the coefficient formulae, and we leave the extension to $n = 10$ for future work.
Furthermore, our framework is also adaptable to complex scheme parameters, which have not been explored much outside of some schemes at order $n = 4$~\cite{Ostmeyer:2022}.
Yet another avenue of exploration is the specialisation of the general Trotter-Suzuki framework to symplectic integrators, where the condition $[B,[B,[B,A]]] = 0$ reduces the commutator basis.
Construction of symplectic integrators could be beneficial to classical simulations, such as molecular or astronomical dynamics.
Work on this adaptation to our framework is ongoing.
Finally, application of high-order schemes to other physical systems could be of great interest to certain fields.
For example, Trotter error bounds for lattice field theories in the Hamiltonian formalism could provide improved resource cost estimates for simulations on quantum hardware.

\section*{Code and Data}

The code and scheme parameters needed to reproduce the results in this manuscript are available on GitHub and Zenodo~\cite{markomalezic_2026_18347430}.
The recursive formulae were originally derived with Wolfram Mathematica~\cite{Mathematica}, using routines provided by FeynCalc~\cite{MERTIG1991345, Shtabovenko2025}.
We also ported the derivation to the open-source Python package \texttt{pybch}, based on SymPy~\cite{sympy}, which is included in the same repository.
The minimisation framework is written in \texttt{C++}, using the linear algebra library Eigen~\cite{eigenweb}.
The noiseless time evolution from \Cref{sec:numerical_performance} is implemented in \texttt{C++}, and the noisy circuit analysis from \Cref{sec:quantum_circuits} in Python.
We constructed the circuits with Qiskit~\cite{qiskit2024}, using our plugin \texttt{qiskit-omelyan}~\cite{qiskit-omelyan} to synthesise the time evolution operator into Pauli rotation gates.
The noisy circuits were simulated with Qiskit Aer~\cite{qiskit2024}, and zero-noise extrapolation was performed with Mitiq~\cite{LaRose2022mitiqsoftware}.
The data from our time evolution simulations can be reproduced with the provided code, and are available from the authors upon request.

\begin{acknowledgments}
	We would like to thank
	Luca Dellantonio, Martina Gisti, Emanuele Mendicelli, and Jinzhao Sun for insightful discussions and literature recommendations.
  The authors gratefully acknowledge the access to the Marvin cluster of the University of Bonn.
  We acknowledge the use of Claude in proofreading of this manuscript, as well as software debugging.
  This work was funded by the Deutsche Forschungsgemeinschaft (DFG, German Research Foundation) as part of the CRC 1639 NuMeriQS – Project number 511713970.
\end{acknowledgments}

\bibliography{bibliography}

\appendix


\section{Additional scheme parameters}\label{app:additional_schemes}

\begin{figure*}[t]
	\centering
	\resizebox{0.85\textwidth}{!}{%
		\input{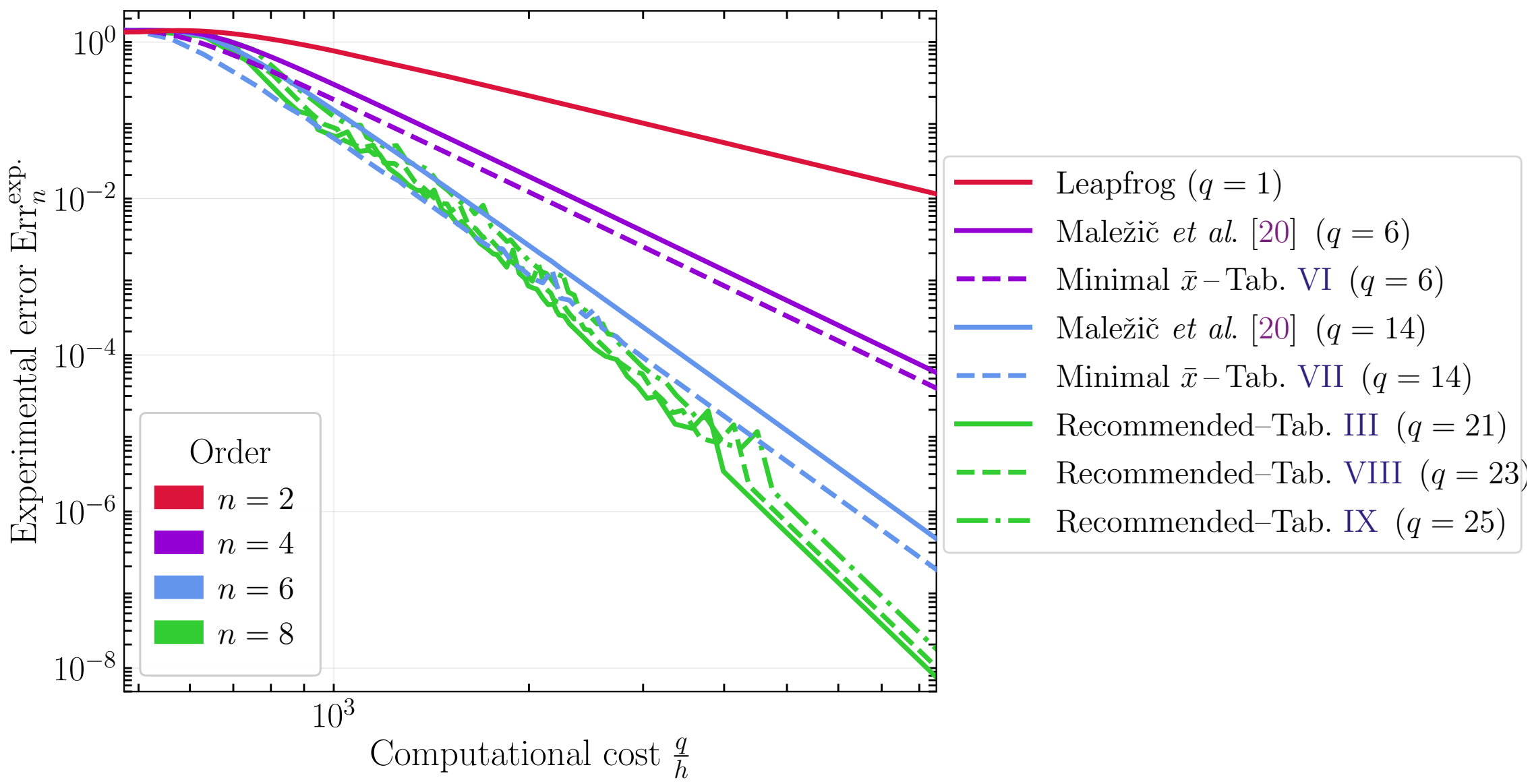}
	}
	\caption{Experimental error $\textrm{Err}_{n}^{\textrm{exp.}}$~\eqref{eq:exp_error} with respect to the computational cost $\nicefrac{q}{h}$ needed to compute the time evolution of a Heisenberg XZ spin chain with $L = 6$ sites and open boundary conditions.
           We plot a selection of additional schemes in the local grouping $S^{(2L)}$ at a fixed total time $t = 100$ and compare them to the Leapfrog scheme, our recommendations in~\cite{Malezic_2026} and our $q = 21$ scheme (\Cref{tab:scheme_q21}).
           The minimal $\bar{x}$ schemes at lower orders $n = 4, 6$ reduce the error slightly in this scenario.
           While the additional schemes at $8^{\textrm{th}}$ order do not improve on our recommended scheme at $q=21$ cycles, they remain competitive.}\label{fig:additional_schemes}
\end{figure*}

During our numerical tests, we noticed that some schemes performed well, although they were not the most efficient at their order.
These schemes favour more uniform parameters, with some of the smallest distances from the origin $\bar{x}$ that we found.
Here we give such schemes at orders $n = 4, 6, 8$ as additional recommendations, and compare their practical performance on the Heisenberg XZ model to our main recommendations in \Cref{fig:additional_schemes}.

Firstly, we present two schemes at lower orders $n = 4, 6$.
Both are at the maximal number of cycles $q = 6$ and $q = 14$, and have the minimal distance from the origin $\bar{x}$ at their $q$.
Their parameters can be found in Tables~\ref{tab:scheme_q6_minimal} and~\ref{tab:scheme_q14_minimal} respectively.
Simulating the Heisenberg model in the local grouping, we find they slightly improve on our recommended schemes~\cite{Malezic_2026} in the accumulated error $\textrm{Err}_{n}^{\textrm{exp.}}$ (see Fig.~\ref{fig:additional_schemes}).
This supports our suggestion in \Cref{sec:numerical_performance} that a stronger preference for uniformity can be beneficial in the local ordering.

\begin{table}[h!]
  \centering
	\caption{Scheme parameters $c_{j}$ for a $4^{\textrm{th}}$ order scheme at maximal cycles $q = 6$.
           The scheme has the minimal distance from the origin $\bar{x}$ at this $q$, which comes at the cost of a lower efficiency.}\label{tab:scheme_q6_minimal}
  \begin{tabular}{C{0.5cm}S[table-format=-1.18, group-digits=none, table-column-width=5.75cm]}
		\toprule
		\multicolumn{2}{c}{$\textrm{Eff}_{4} = \num[round-mode=figures,round-precision=4]{4.829333}$} \\
		\multicolumn{2}{c}{$\bar{x} = \num[round-mode=figures,round-precision=3]{0.662755}$} \\
		\midrule
		$j$ &                   {$c_{j} = d_{q+1-j}$}   \\
		\midrule
		1 &  0.287834613619764812 \\
		2 &  0.151098631207585238 \\
		3 &  0.154721182891605497 \\
		4 &  0.112757707574057481 \\
		5 &  0.117761677709795246 \\
		6 & -0.324173813002808231 \\
		\bottomrule
	\end{tabular}
\end{table}

\begin{table}[h!]
  \centering
	\caption{Scheme parameters $c_{j}$ for a $6^{\textrm{th}}$ order scheme at maximal cycles $q = 14$.
           The scheme has the minimal distance from the origin $\bar{x}$ at this $q$, which comes at the cost of a lower efficiency.}\label{tab:scheme_q14_minimal}
  \begin{tabular}{C{0.5cm}S[table-format=-1.18, group-digits=none, table-column-width=5.75cm]}
		\toprule
		\multicolumn{2}{c}{$\textrm{Eff}_{6} = \num[round-mode=figures,round-precision=3]{0.150275}$} \\
		\multicolumn{2}{c}{$\bar{x} = \num[round-mode=figures,round-precision=3]{0.637057}$} \\
		\midrule
		$j$ &                   {$c_{j} = d_{q+1-j}$}   \\
		\midrule
		1  &  0.043704135948876253 \\
		2  &  0.085971959351543434 \\
		3  & -0.140210973528653471 \\
		4  &  0.235739169767919005 \\
		5  &  0.098175286900802475 \\
		6  &  0.094535865815963044 \\
		7  &  0.078981222193509087 \\
		8  &  0.068583665915761011 \\
		9  &  0.082485694645635016 \\
		10 & -0.165420520751213745 \\
		11 & -0.226675480808975371 \\
		12 &  0.057382351788547323 \\
		13 &  0.108345004651375282 \\
		14 &  0.078402618108910652 \\
		\bottomrule
	\end{tabular}
\end{table}

At $8^{\textrm{th}}$ order, we give two additional compositions found with the adapted framework (see \Cref{sec:optim_improve}).
The first one has $q = 23$ (see Tab.~\ref{tab:scheme_q23}) with the highest efficiency at this $q$, and slightly smaller distance from the origin $\bar{x}$ than our recommendation at $q = 21$ cycles.
It is also slightly less efficient, but these properties could make it a generally well performing scheme.
The other composition has $q = 25$ cycles (see Tab.~\ref{tab:scheme_q25}), and it balances efficiency and uniformity, with a distance $\bar{x}$ comparable to our recommendation.
We benchmark both against our recommendation (see Fig.~\ref{fig:additional_schemes}), and find that they fall slightly short.
Nonetheless, they are competitive and could perform better in other models.

\begin{table}[h!]
  \centering
	\caption{Scheme parameters $c_{j}$ for an $8^{\textrm{th}}$ order composition scheme at $q = 23$.
           The composition is slightly more uniform than our recommended scheme at $q = 21$ cycles, at roughly 70 \% of its efficiency.
           Due to the mirror symmetry $c_{j} = c_{q+1-j}$ of composition schemes (see Sec.~\ref{sec:manifold_structure}), we can compress the table to half the size.}\label{tab:scheme_q23}
  \begin{tabular}{C{1.0cm}S[table-format=-1.18, group-digits=none, table-column-width=5.75cm]}
		\toprule
		\multicolumn{2}{c}{$\textrm{Eff}_{8} = \num[round-mode=figures,round-precision=4]{4.32526504e-4}$} \\
		\multicolumn{2}{c}{$\bar{x} = \num[round-mode=figures,round-precision=3]{0.66111873}$} \\
		\midrule
		$j$ &                   {$c_{j} = d_{q+1-j}$}   \\
		\midrule
		1, 23  &  0.035775813126863683 \\
		2, 22  &  0.082428855761252390 \\
		3, 21  & -0.093558098054503130 \\
		4, 20  &  0.044803882268296111 \\
		5, 19  &  0.041609234548577574 \\
		6, 18  &  0.037205238783110939 \\
		7, 17  &  0.039574307245707777 \\
		8, 16  &  0.194301912776415037 \\
		9, 15  & -0.151953430890394953 \\
		10, 14 &  0.040812082768358217 \\
		11, 13 &  0.083271171704635674 \\
		12     & -0.208541940076638649 \\
		\bottomrule
	\end{tabular}
\end{table}

\begin{table}[h!]
  \centering
	\caption{Scheme parameters $c_{j}$ for an $8^{\textrm{th}}$ order composition scheme at $q = 25$.
           The scheme is less than half as efficient as our recommended scheme at $q = 21$, with a comparable distance $\bar{x}$.
           Due to the mirror symmetry $c_{j} = c_{q+1-j}$ of composition schemes (see Sec.~\ref{sec:manifold_structure}), we can compress the table to half the size.}\label{tab:scheme_q25}
  \begin{tabular}{C{1.0cm}S[table-format=-1.18, group-digits=none, table-column-width=5.75cm]}
		\toprule
		\multicolumn{2}{c}{$\textrm{Eff}_{8} = \num[round-mode=figures,round-precision=4]{2.63551680e-4}$} \\
		\multicolumn{2}{c}{$\bar{x} = \num[round-mode=figures,round-precision=3]{0.68798272}$} \\
		\midrule
		$j$ & {$c_{j} = d_{q+1-j}$}\\
		\midrule
		1, 25  &  0.047951181370257979 \\
		2, 24  &  0.058306285798459791 \\
		3, 23  &  0.081833806671706782 \\
		4, 22  & -0.099767039618495923 \\
		5, 21  & -0.081108207667039356 \\
		6, 20  &  0.056506301170499422 \\
		7, 19  &  0.061992066674556361 \\
		8, 18  &  0.069082895721143928 \\
		9, 17  &  0.188059120467134577 \\
		10, 16 & -0.142948183144266994 \\
		11, 15 &  0.050775495147567268 \\
		12, 14 &  0.060649773832167270 \\
		13     & -0.202666992847382155 \\
		\bottomrule
	\end{tabular}
\end{table}

\end{document}